\documentclass[sigconf]{acmart}

\AtBeginDocument{%
  \providecommand\BibTeX{{%
    \normalfont B\kern-0.5em{\scshape i\kern-0.25em b}\kern-0.8em\TeX}}}

\copyrightyear{2026}
\acmYear{2026}
\setcopyright{cc}
\setcctype{by}
\acmConference[IDC '26]{Proceedings of the 25th Interaction Design and Children Conference}{June 22--25, 2026}{Brighton, United Kingdom}
\acmBooktitle{Proceedings of the 25th Interaction Design and Children Conference (IDC '26), June 22--25, 2026, Brighton, United Kingdom}
\acmDOI{10.1145/3773077.3806144}
\acmISBN{979-8-4007-2283-7/2026/06}

\usepackage{graphicx}  

\usepackage{amssymb}   
\usepackage{colortbl}
\usepackage{array}
\usepackage{caption}
\usepackage{subcaption}
\usepackage{geometry}
\usepackage{xcolor} 
\usepackage{booktabs}
\usepackage{multirow}
\usepackage{tabularx}
\usepackage{graphicx}
\usepackage{longtable}
\usepackage{appendix}
\usepackage{pifont}

\begin{document}

\author{Aayushi Dangol}
\affiliation{%
  \institution{University of Washington}
  \city{Seattle}
  \state{WA}
  \country{USA}
}
\email{adango@uw.edu}

\author{Robert Wolfe$^*$}
\thanks{$^*$Both authors contributed equally.}
\affiliation{%
  \institution{Rutgers University}
  \city{New Brunswick}
  \state{NJ}
  \country{USA}
}
\email{robert.wolfe@rutgers.edu }

\author{Nisha Devasia$^*$}
\affiliation{%
  \institution{University of Washington}
  \city{Seattle}
  \state{WA}
  \country{USA}
}
\email{ndevasia@uw.edu}

\author{Mitsuka Kiyohara}
\affiliation{%
  \institution{ University of Toronto}
  \city{Toronto}
  \state{Ontario}
  \country{Canada}
}
\email{ndevasia@uw.edu}

\author{Jason Yip}
\affiliation{%
  \institution{University of Washington}
  \city{Seattle}
  \state{WA}
  \country{USA}
}
\email{jcyip@uw.edu}

\author{Julie A. Kientz}
\affiliation{%
  \institution{University of Washington}
  \city{Seattle}
  \state{WA}
  \country{USA}
}
\email{jkientz@uw.edu}

\title{Where Does AI Leave a Footprint? Children’s Reasoning About AI’s Environmental Costs}

\begin{abstract}

Two of the most socially consequential issues facing today’s children are the rise of artificial intelligence (AI) and the rapid changes to the earth’s climate. Both issues are complex and contested, and they are linked through the notable environmental costs of AI use. Using a systems thinking framework, we developed an interactive system called Ecoprompt to help children reason about the environmental impact of AI. EcoPrompt combines a prompt-level environmental footprint calculator with a simulation game that challenges players to reason about the impact of AI use on natural resources that the player manages. We evaluated the system through two participatory design sessions with 16 children ages 6–12. Our findings surfaced children’s perspectives on societal and environmental tradeoffs of AI use, as well as their sense of agency and responsibility. Taken together, these findings suggest opportunities for broadening AI literacy to include systems-level reasoning about AI’s environmental impact.
\end{abstract}



\renewcommand{\shortauthors}{Dangol et al.}
\maketitle

\section{Introduction}

The rapid rise of generative AI (genAI) has renewed public attention to the environmental costs of large-scale computation. Across both media coverage and academic work, genAI is increasingly recognized as resource-intensive, with substantial energy, water, and material demands associated with training and operating large models \cite{inie2025co2stly,jegham2025hungry}. Recent estimates help ground these concerns. For example, Jegham et al. \cite{jegham2025hungry} report that the 700 million daily queries made to the GPT-4o model alone required ``electricity use comparable to 35,000 U.S. homes, freshwater evaporation matching the annual drinking needs of 1.2 million people, and carbon emissions requiring a Chicago-sized forest to offset.''

Concerns about the ecological footprint of computation are not new. Other data-intensive technologies, such as cryptocurrency mining, cloud gaming, and large-scale data storage, have long relied on energy-hungry data centers and globally distributed infrastructures \citep{hans2014did, moutaib2020internet, li2019energy}. What distinguishes genAI, however, is not only the scale of its infrastructure, but the breadth and immediacy of its adoption by the public. Since late 2022, genAI tools have moved rapidly from specialized applications into everyday use, embedded in search engines, productivity tools, educational platforms, and consumer-facing applications \citep{pan2025complexity, noy2023experimental, suh2024opportunities, lewis2025exploring}. This expansion has dramatically increased the number of people who interact directly with genAI systems. For example, ChatGPT alone grew from about 1 million users within five days of its launch in November 2022 to over 800 million weekly active users by late 2025, reaching roughly 10\% of the world’s adult population \cite{bellan2025users, perez2025growth}. 

Despite the accumulating evidence of genAI’s environmental impact, these costs remain largely invisible to end users. At the same time, genAI is increasingly embedded in the tools children use for learning, creativity, and play \citep{Dangol2025, zhang2021storydrawer, wu2023integrating}, bringing a growing population of young users into routine contact with AI systems without clear information or opportunities to engage with those impacts. This invisibility is especially consequential given that prior work shows that children are already attentive to climate change and environmental sustainability, often expressing concern about environmental futures and a desire to act responsibly \citep{burke2018psychological, wu2015climate, lee2020youth}. Yet far less is known about how children perceive, interpret, and reason about the environmental impacts of AI systems. This gap motivates us to ask the following research questions:

\begin{itemize}
\item[] \textbf{RQ1:} How do children (ages 6 - 12) understand and reason about the environmental costs of genAI systems?
\item[] \textbf{RQ2:} What are some best practices for designing interactive systems that make the ecological footprint of genAI visible, relatable, and actionable for children?
\end{itemize}

Recognizing that children are often drawn to games and simulations as ways of exploring complex ideas \cite{peppler2013collaborative, devasia2020escape, divanji2024togethertales}, we developed EcoPrompt, a web-based system that invites children to explore how everyday interactions with genAI are connected to underlying environmental costs. In designing EcoPrompt, we drew on prior work in systems thinking \citep{arnold2015definition, green2021empirical}, which emphasizes helping learners understand how individual actions are connected to broader systems and how their effects accumulate over time \citep{dangolreading}. We then conducted two participatory design sessions with 16 children (ages 6–12).  Our findings show that as children engaged with EcoPrompt they developed increasingly systems-oriented explanations that connected genAI use to broader infrastructures such as electricity, data centers, and water-intensive cooling. Making genAI’s resource use legible also prompted children to set limits on their genAI use, debate whether particular questions were worth their environmental cost, and evaluate genAI use in terms of its informational return relative to resource expenditure. Through these discussions, children began negotiating responsibility for shared resources (e.g., energy and water) and critiquing forms of genAI use they viewed as wasteful or unproductive.

Our contributions offer both design insights and an empirical understanding of how children make sense of genAI’s environmental impact and develop strategies for deciding when to use or refrain from using genAI.  In the remainder of this paper, we begin by reviewing related work. We then describe our study’s methodological approach, including our data collection, analysis procedures, and the design of EcoPrompt. Finally, we present our findings and discuss their implications for supporting children’s AI literacy.

\section{Background \& Related Work}

\subsection{Environmental Impacts of Generative AI}
The rapid expansion of large language models (LLMs) and other genAI systems has been accompanied by unprecedented growth in computational infrastructure \citep{tan2026sky, satariano2026microsoftwater}. GenAI systems rely on large-scale data centers equipped with specialized hardware, such as GPUs and accelerators, which require significant amounts of electricity to support model training and ongoing operation \citep{cottier2024rising}. It is notable that other energy-intensive technologies of the past decade, such as cryptocurrency and cloud gaming \citep{hans2014did, moutaib2020internet, li2019energy}, have not made the same impact on the technology industry's energy consumption as AI has; from 2005 to 2017, the amount of electricity going to data centers was relatively unchanging, as increases in efficiency mitigated energy consumption resulting from increases in demand \cite{O’Donnell_Crownhart_2025}. However, recent investments in data centers from both governments and industry at large have increased the strain that data centers place on the power grid; recent estimates state that data centers use about 4.4\% of the U.S. domestic electricity supply, with the number expected to double in the next few years \cite{Energy.gov_2024}. Furthermore, in many regions, this electricity is still predominantly generated through fossil fuel based sources, including coal and natural gas, linking genAI use to greenhouse gas emissions and local air pollution \citep{iea2023datacenters, patterson2021carbon}. Beyond electricity consumption, genAI systems place considerable demands on freshwater resources \citep{satariano2026microsoftwater}. Data centers generate substantial heat during computation and require continuous cooling to maintain reliable operation. In many facilities, cooling is achieved through water-intensive systems, leading to significant freshwater withdrawals \citep{luccioni2023counting}. 

These impacts are highly context-dependent and can be particularly consequential in regions already experiencing water stress or prolonged drought. Recent estimates suggest that AI-related data center operations may consume billions of gallons of freshwater annually, raising concerns about the sustainability of continued infrastructure expansion in water-scarce areas \citep{li2025makingaithirstyuncovering, un2025water}. The environmental footprint of genAI also extends upstream to the production of hardware and supporting infrastructure. The manufacture of GPUs, servers, and networking equipment requires the extraction and processing of critical minerals and rare earth elements, including copper, cobalt, and lithium, which are often sourced through environmentally destructive mining practices \citep{ligozat2022unraveling}. Hardware production is energy-intensive and contributes to emissions and material depletion well before a model is deployed, while rapid hardware obsolescence generates large volumes of electronic waste that are frequently inadequately recycled \citep{un2025water}.

While training a model is widely recognized as computationally intensive \citep{O’Donnell_Crownhart_2025}, prior research emphasizes that the cumulative environmental footprint of genAI is often shaped by the generation of outputs in response to user prompts, as repeated, everyday interactions scale energy and water use across a model’s lifetime \citep{luccioni2023counting}. Despite growing awareness of these issues within the AI and sustainability communities, the environmental costs of genAI remain largely invisible to end users, particularly children. Unlike other energy-intensive technologies, children have easy access to genAI systems, which are becoming increasingly popular in educational, creative, and everyday contexts \citep{commonsense2025ai}. Without support for understanding how genAI systems are produced, powered and sustained users may reasonably interpret AI interactions as immaterial or consequence-free \cite{inclezan2025environment}. Foregrounding the environmental dimensions of genAI is therefore essential not only for supporting informed and responsible use \citep{dangol2025beyond}, but also for enabling children to situate AI within broader societal conversations about sustainability, equity, and the environmental costs of digital technologies.

\subsection{AI Literacy, Societal Impact, and Generative AI}

In the past decade, AI-enabled technologies have become increasingly embedded in children’s everyday lives, including interactive toys \cite{kewalramani2021using, williams2018my}, voice assistants \cite{druga2017hey}, social robots \cite{williams2019artificial}, and, more recently, genAI systems such as ChatGPT \cite{newman2024want, ali2024constructing}. As these technologies have proliferated, researchers and educators have articulated frameworks and guidelines for K–12 AI literacy, broadly defined as the knowledge, skills, and perspectives needed for learners to critically evaluate and meaningfully engage with AI technologies \cite{long2020ai, touretzky2019envisioning, zhang2023integrating, lee2021developing}. One such AI literacy framework is the “Five Big Ideas of AI,” which includes societal impact as a core dimension, emphasizing the potential positive and negative consequences of AI systems for individuals, communities, and broader social structures \cite{touretzky2019envisioning}.

In existing AI literacy research, the societal impact big idea has most often been taken up through attention to issues of algorithmic fairness and bias. For example, prior work has supported children in examining gender bias in supervised machine learning, foregrounding the role of training data, and exploring how biased datasets shape system outputs \cite{williams2023ai+, payne2019ethics}. Others have used games \cite{dangol2025ai}, participatory design activities \cite{dangol2024mediating}, and constructionist approaches \cite{jordan2021poseblocks, druga2018growing} to integrate technical AI concepts with discussions of ethics and social responsibility. Together, this body of work has been instrumental in demonstrating that children are capable of engaging with AI not only as users, but as critical thinkers attentive to issues of harm, responsibility, and power.

Building on prior work, we argue that the societal impact big idea can also be extended to include the environmental impact of genAI. Increases in general digital literacy are thought to support the adoption of eco-friendly behaviors such as recycling, energy conservation, and reduction of single-use plastics \cite{chen2025exploring}. However, AI consumer usage and large industry investments in increased scaling are exacerbating ecosystem degradation and resource depletion at an unprecedented scale that existing eco-literacy frameworks do not address \cite{inclezan2025environment}. 

It is crucial for children to understand AI's multifacted impacts to become conscientious consumers of the technology \cite{payne2019ethics, lee2021developing, long2020ai, dipaola2022preparing}. As AI literacy seeks to support learners in reasoning about how AI systems shape the world around them, attending to material and ecological aspects of genAI, such as energy use, water consumption, and infrastructure demands, offers an additional dimension for reflection. Prior scholarship in CCI and sustainability education provides design precedents for supporting this kind of reasoning, emphasizing children’s engagement with technologies as embedded within interconnected social, material, and ecological systems rather than as isolated tools \citep{wilensky1999thinking, resnick1998diving, green2021empirical}. Drawing on this tradition, in the next section we introduce systems thinking as a framework for helping children reason about the environmental impact of genAI.

\subsection{Systems Thinking in Sustainability and Environmental Education}
Systems thinking is a framework that supports reasoning about complex phenomena by foregrounding relationships among interconnected components, resource flows, and feedback processes that unfold over time and across scales \citep{arnold2015definition, meadows2008thinking, wiek2011key}. Within sustainability and environmental education, systems thinking has been widely adopted to help learners understand environmental challenges such as climate change, resource depletion, and pollution as outcomes of interacting social, material, and ecological systems \citep{green2021empirical, peretz2025integrating, blatti2019systems}. Prior research emphasizes the importance of supporting children’s ability to shift between local and global viewpoints, enabling them to connect everyday actions and practices to distributed, cumulative, system-level consequences \citep{green2021empirical, church2010sustainability}. Scholars further argue that encouraging reflection on how children situate themselves within these systems, shaped by social, cultural, and intersectional contexts, can deepen engagement and support more meaningful forms of understanding \citep{kaijser2014climate}.

Within climate and sustainability education, systems thinking is frequently enacted through interactive simulations and serious games \citep{madani2017serious, robinson1983game, novaes2025enhancing}, which allow children to explore how individual actions interact within shared systems in engaging ways. Early sustainability-focused serious games often took the form of tabletop modalities such as card and board games \cite{robinson1983game, eisenack2013climate, antle2014emergent}, with digital games becoming increasingly prevalent in more recent work \cite{reckien2013climate}. Across both formats, sustainability-focused serious games typically employ role-play or management simulation mechanics \cite{reckien2013climate, shapiro2011games} and commonly address themes such as resource management, irrigation, carbon emissions, and deforestation \citep{madani2017serious, barnes2017exploring, wise2015kind}. Through repeated interaction, such games can support systems thinking by enabling children to observe how individual decisions accumulate into collective outcomes, often producing unintended or emergent effects \citep{wise2021design}.

A systems thinking lens is also well-suited to the case of genAI, as its environmental impact is largely invisible to end users and distributed across distant infrastructures such as data centers, energy grids, and water systems. As a result, children’s reasoning about genAI’s environmental footprint requires bridging everyday interactions with abstract, system-level processes that unfold across space and time. Given the growing role of genAI in children’s everyday learning and creative practices \citep{dangol2024ai, dangol2025if, pew2025chatgpt}, this raises an important question: how might we support children in connecting their own genAI use to these distributed environmental impacts and reason across both individual and collective scales? Guided by prior work on systems thinking and AI literacy, the next section introduces EcoPrompt, an interactive system designed to support children’s reasoning about genAI’s environmental footprint.

\section{EcoPrompt: System Design \& Development}
\begin{figure}[ht]
    \centering
    \includegraphics[width=1\linewidth]{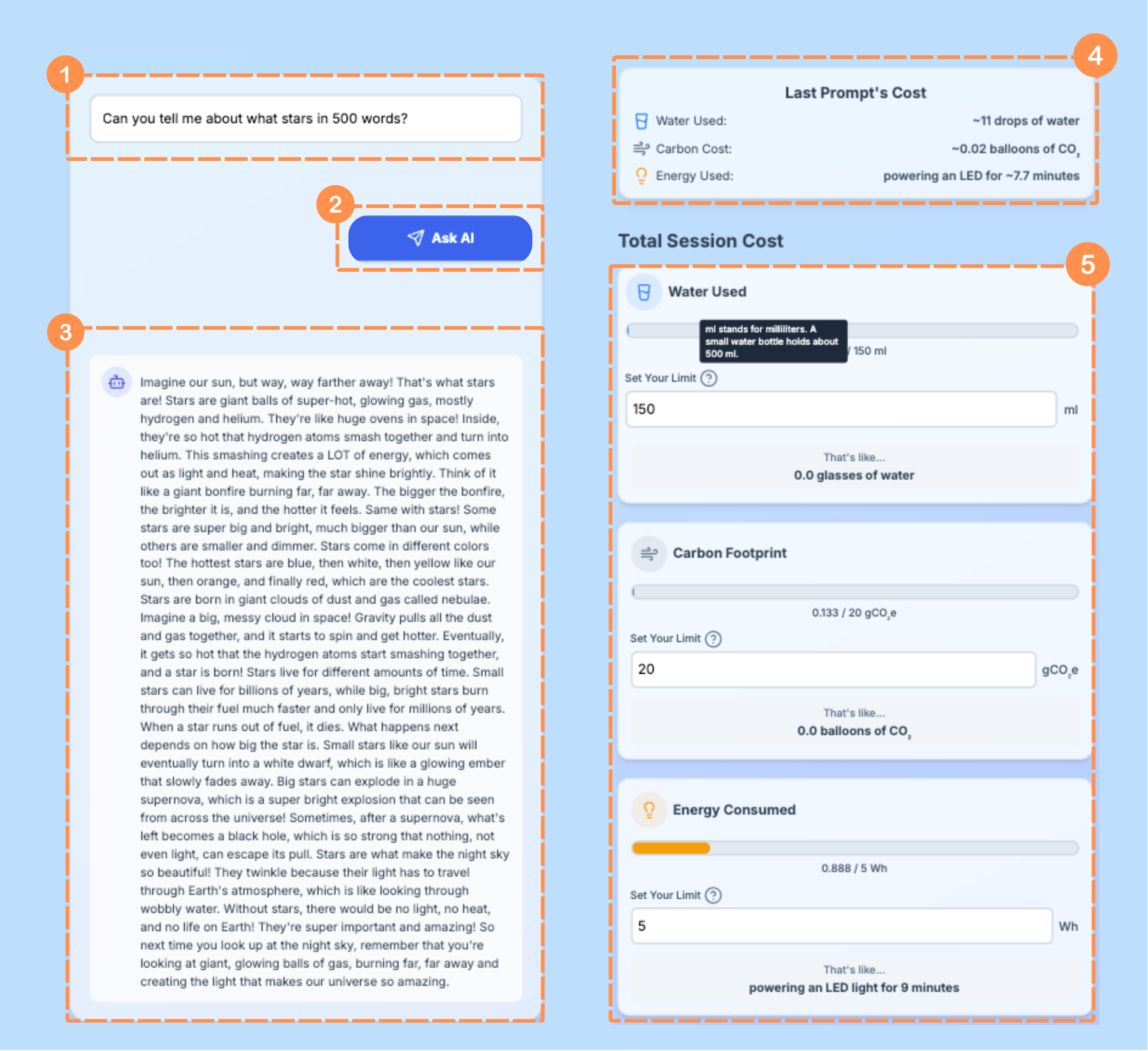}
    \caption{Footprint Calculator interface, including (1) user prompt entry box, (2) an "Ask AI" button that submits the prompt to the model, (3) an output box in which the AI's response is displayed, (4) a display of the last prompt's estimated water, carbon, and energy costs, and (5) a display of the total water, carbon, and energy costs across the session, with status-bar style indicators of cumulative use.}
    \label{fig:calculator}
\end{figure}

\begin{figure}[ht]
    \centering
    \includegraphics[width=1\linewidth]{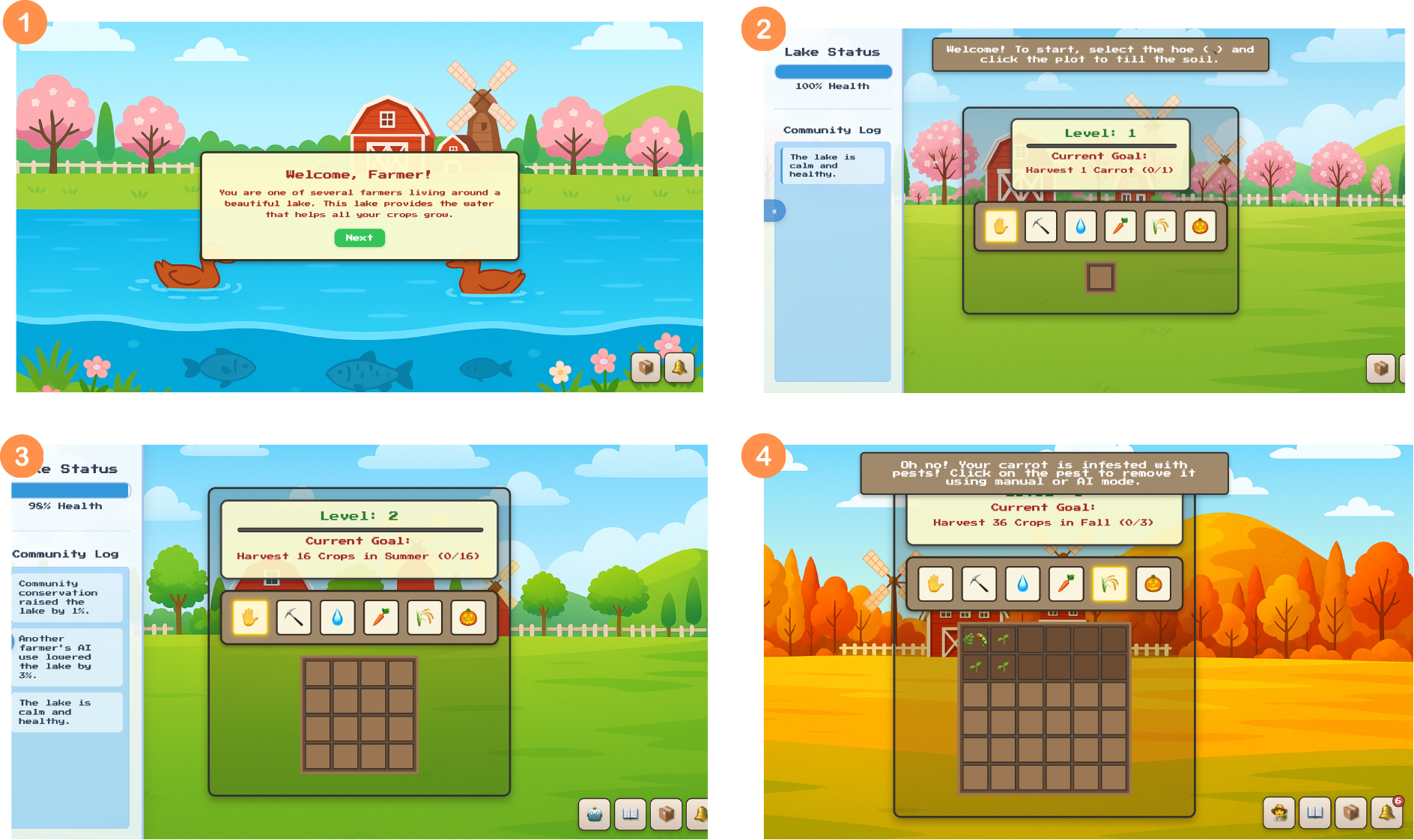}
    \caption{Farm game interface showing (1) initial welcome screen, (2-4) gameplay across three levels. In each level, players face various challenges. For example, in (4), pest attacks introduce critical decision points where players can use AI to make crop protection more manageable as levels progress.}
    \label{fig:crops}
\end{figure}

EcoPrompt is an interactive system designed to support children in reasoning about the environmental impacts of genAI. Children interact with EcoPrompt through two complementary but separate interfaces: (1) a Footprint Calculator, which provides real-time estimates of energy, water, and carbon use associated with children’s genAI prompts (see Figure \ref{fig:calculator}); and (2) a Farm Game, in which children manage virtual farms and can choose whether and when to use genAI to support their farming activities (see Figure \ref{fig:crops}). While the Footprint Calculator foregrounds prompt-level resource use and individual decision-making, the Farm Game supports reflection on collective dynamics, accumulation, and longer-term environmental consequences. Together, these interfaces broaden children’s inquiry from “What resources does this single genAI interaction use?” to “How do patterns of genAI use interact with shared ecological systems?” Below, we describe the design considerations that guided EcoPrompt’s development, followed by an overview of its two core components.

\subsection{Design Considerations}

\subsubsection{Making GenAI's Environmental Impacts Visible through Prompt-Level Feedback} Prior work in systems thinking and AI literacy emphasizes that learners often struggle to reason about complex systems when key processes remain invisible or abstract \citep{green2021empirical, dangol2025children, williams2023ai+}. EcoPrompt addresses this challenge by making the environmental impacts of genAI visible at the moment of interaction through the Footprint Calculator. By providing real-time estimates of energy, water, and carbon use associated with children’s genAI prompts, the system foregrounds otherwise hidden infrastructural costs of AI use. This immediate feedback supports children in attending to specific system elements and their relationships, encouraging reflection on how different prompts are associated with different environmental implications. 

\subsubsection{Scaffolding Systems Thinking across Individual and Collective Perspectives}
Understanding environmental systems requires attention not only to individual actions, but also to how effects accumulate over time within systems shaped by multiple actors \citep{blatti2019systems, church2010sustainability}. For children, this means relating their own genAI use not only to shared environmental resources, but also to the actions and constraints of others operating within the same system. While the Footprint Calculator foregrounds individual prompt-level decision-making, the Farm Game situates genAI use within a shared environment where outcomes emerge from the interplay of many actors, not all of whom are under the child’s control. This shifts attention away from immediate, isolated cause-and-effect relationships toward longer-term and collective consequences that unfold even when children make careful or restrained choices. In doing so, the Farm Game supports reasoning about responsibility and impact in contexts where agency is distributed and outcomes are uncertain. By separating these perspectives across complementary interfaces, EcoPrompt allows children to engage with both the individual and collective dynamics of genAI use without requiring them to reason about all aspects simultaneously.

\subsubsection{Encouraging Deliberation and Trade-Off Reasoning in Decisions about GenAI Use}
Prior work in HCI has characterized decision-making around genAI use as a wicked problem, emphasizing that such decisions involve competing values and no single correct solution \citep{corbin2025wicked, moulaison2025wicked}. Therefore, EcoPrompt is designed to foreground reflection, and trade-off reasoning across multiple levels of decision-making. At the prompt level, the Footprint Calculator supports deliberation as children can set limits on their genAI use and receive feedback when those limits are approached or exceeded, creating opportunities to pause, reconsider, and renegotiate their choices. The Farm Game extends this deliberative framing to a broader, collective context. Within the game, children are not required to use genAI, but can choose whether and when it is appropriate to do so in support of shared goals, positioning genAI as one option among many within a dynamic system. Together, these mechanisms promote ongoing deliberation rather than one-time judgments, supporting children in reasoning about genAI use as a situated and evolving decision shaped by both individual actions and collective outcomes.

\subsubsection{Choice to Use a Real Generative AI API} We faced the difficult choice of whether to create a system that 1) connected to a real generative AI system, dynamically displaying outputs to the player, or 2) attempted to mimic the output of such a system without actually calling the API. The former option would provide children with a more faithful and reliable experience, affording the opportunity to interact directly with a generative model while reasoning about its environmental impact, while the latter would avoid the environmental costs associated with using the model. We ultimately chose the former approach, as we wanted children to have the opportunity to experiment with the model and form ideas about how it worked, and we did not believe we could sufficiently replicate this with a more limited system. While this approach did yield the intended results (see for example Section 5.1.3, where children discover that they cannot control the output verbosity of a real generative model and thus easily reduce its environmental footprint), we also acknowledge that the use of a generative model over our two sessions likely resulted in the production of 1-2 kilograms of CO2. We consider this a limitation in our approach, but a necessary one for supporting the ecological validity of the study.

\subsection{System Overview}

\subsubsection{Footprint Calculator} 
\begin{figure}[ht]
    \centering
    \includegraphics[width=1\linewidth]{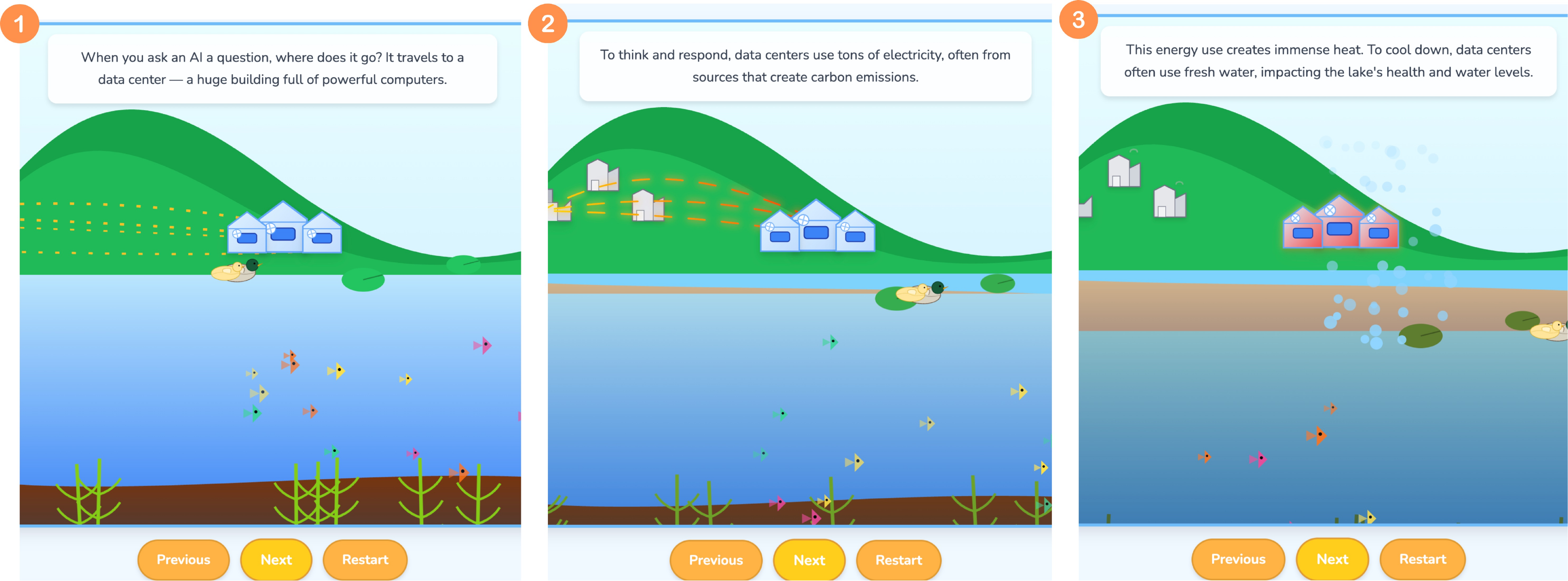}
    \caption{Tutorial illustrating the AI query lifecycle and its environmental consequences in EcoPrompt, demonstrating how a user prompt travels from a personal device to a data center that consumes electricity, resulting in carbon emissions.}
    \label{fig:query-lifecycle}
\end{figure}

The Footprint Calculator begins with a short interactive tutorial that introduces children to the lifecycle of an AI query. The tutorial (see Figure~\ref{fig:query-lifecycle}) depicts how a user’s prompt is transmitted from a personal device to remote data centers, where large-scale computing infrastructure consumes electricity that is often generated from carbon-emitting sources. It then illustrates how this energy use produces heat that must be dissipated through water-intensive cooling processes, with downstream effects on local ecosystems and water availability. 

This walkthrough establishes a conceptual baseline that helps children interpret the calculator’s core indicators (energy, water, and carbon) by making visible how these resources become implicated in a single AI interaction. The calculator presents these indicators as modeled estimates rather than precise measurements, intended to support comparison, reflection, and reasoning about relative environmental impacts rather than exact quantification.

To estimate the environmental footprint associated with each prompt, we used the infrastructure-aware inference benchmarking framework introduced by Jegham et al. \cite{jegham2025hungry}. In this framework, per-query environmental costs are derived by first estimating the electricity consumed during inference based on runtime characteristics (e.g., latency and generation speed), the power draw of GPU and non-GPU system components, and datacenter-level overhead captured through Power Usage Effectiveness (PUE). This per-query energy estimate is then translated into water consumption and carbon emissions using standard environmental multipliers: Water Usage Effectiveness (WUE), which accounts for water used in both on-site cooling and off-site electricity generation, and Carbon Intensity Factors (CIF), which reflect the carbon content of the regional electricity grid. This approach focuses on operational impacts during inference rather than training or hardware manufacturing, allowing environmental costs to be attributed at the level of individual prompts. 

While these calculations are grounded in established modeling approaches, they necessarily simplify complex and variable infrastructures; as such, the resulting values should be interpreted as approximations that make otherwise invisible processes legible to learners. The Footprint Calculator also allows children to set adjustable resource limits; visual indicators fill as these limits are approached. Children can revise these limits dynamically, supporting exploration of trade-offs rather than compliance with fixed thresholds.

\subsubsection{Farm Game}

\begin{figure}[ht]
    \centering
    \includegraphics[width=1\linewidth]{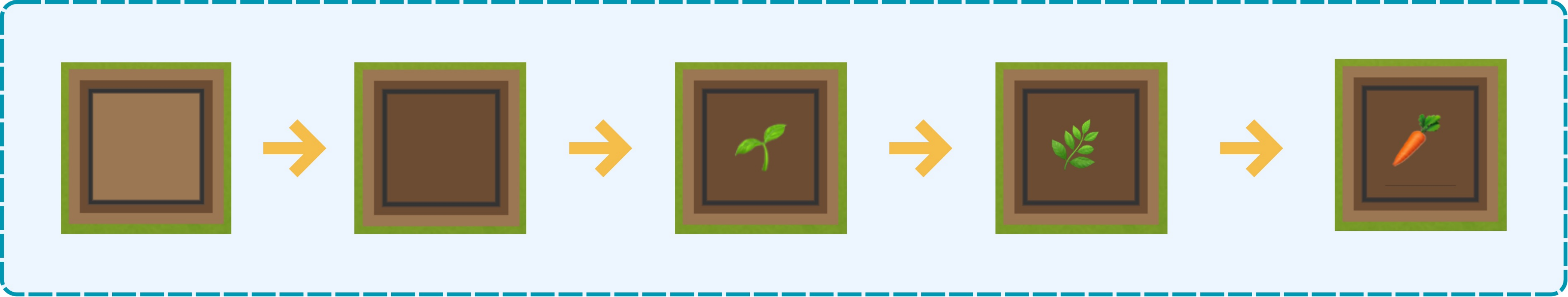}
    \caption{An illustration of crops growing as a result of players' attention to planting, watering, and harvesting them.}
    \label{fig:crops-growing}
\end{figure}

\begin{figure}[ht]
    \centering
    \includegraphics[width=1\linewidth]{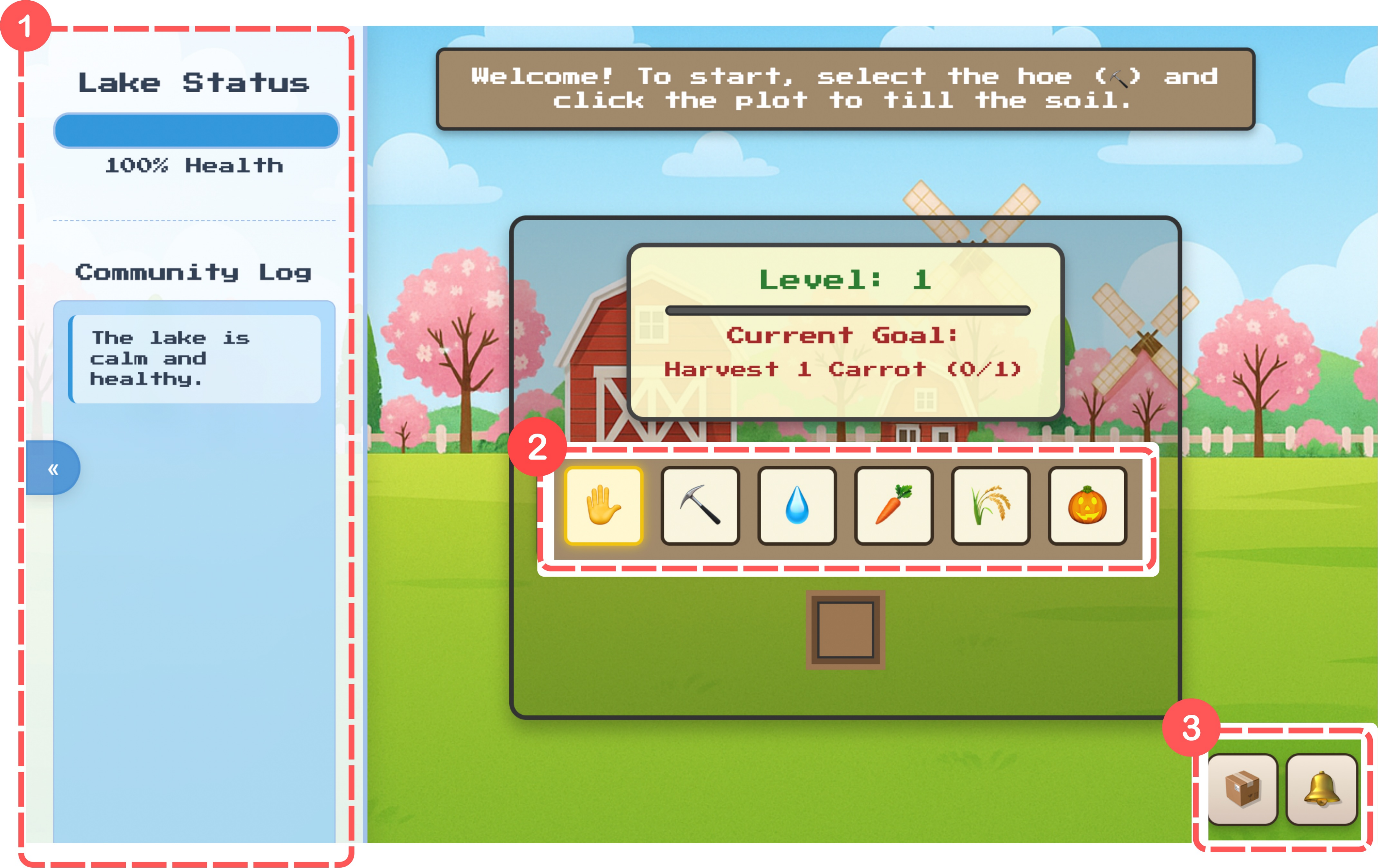}
    \caption{An illustration of the game panel at the start of the first level, showing (1) the sidebar displaying the lake health with a status log communicating changes to the player, (2) the main array of tools available to the player, including the planting and watering tools, and (3) the inventory and notifications panel.}
    \label{fig:community-log}
\end{figure}

The Farm Game is a grid-based farming simulation organized into five levels that progressively introduce new mechanics, challenges, and AI affordances. Players take on the role of a farmer managing crops on a personal farm while drawing from a shared community lake that supplies water and energy to all farms in the region. Across levels, players engage in routine agricultural activities such as planting, watering, harvesting, (see Figure \ref{fig:crops-growing}) and responding to disruptions, while being periodically offered AI-powered assistance. In all levels, the farm is represented as a fixed grid of tiles that players interact with using a toolbar of tools and seeds (see Figure \ref{fig:community-log}). Crops grow over time and contribute to experience points, which unlock new goals and features. Although players cannot see or directly interact with other farmers, the lake’s health reflects the cumulative effects of AI-related actions taken across the system. If the lake’s health reaches zero, players lose the game. Conversely, players who manage their farms successfully can earn more coins and finish the game with the highest score. \newline

\begin{figure}[ht]
\centering
\includegraphics[width=1\linewidth]{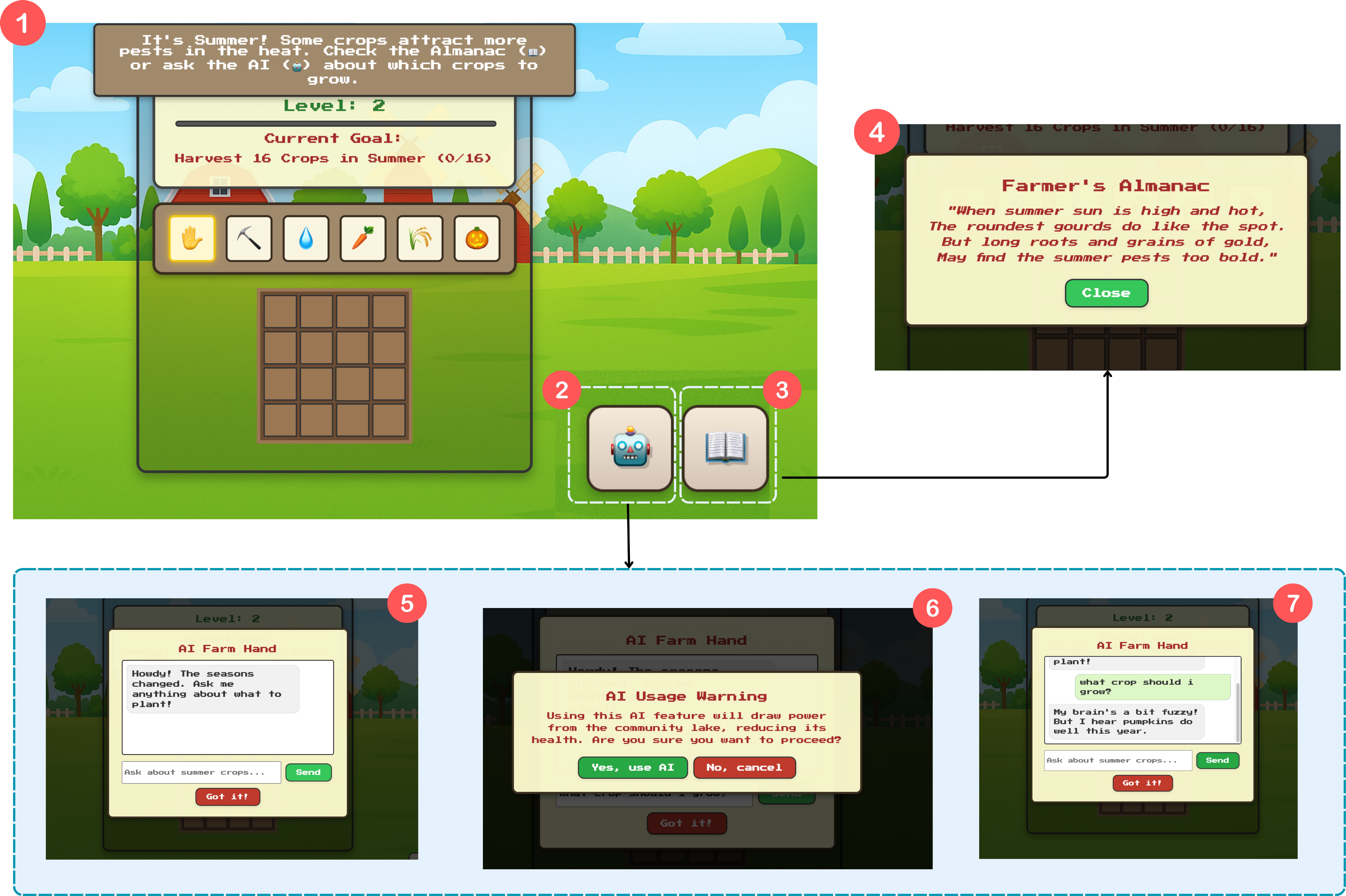}
\caption{The interface for level 2 of the Farm Game, showing (1) the overall UI, (2) the AI Farm Hand button, (3) the Farmer's Almanac button, (4) the Farmer's Almanac hint message displayed after pressing the button, (5) the AI Farm Hand interface, (6) an AI Usage Warning displayed before sending the user's message to the AI Farm Hand, and (7) the response generated by the AI Farm Hand, suggesting what to plant. Levels 3 and 4 retain the Farm Hand and Almanac buttons.}
\label{fig:farm-game}
\end{figure}

\begin{figure} [ht]
\centering
\includegraphics[width=1\linewidth]{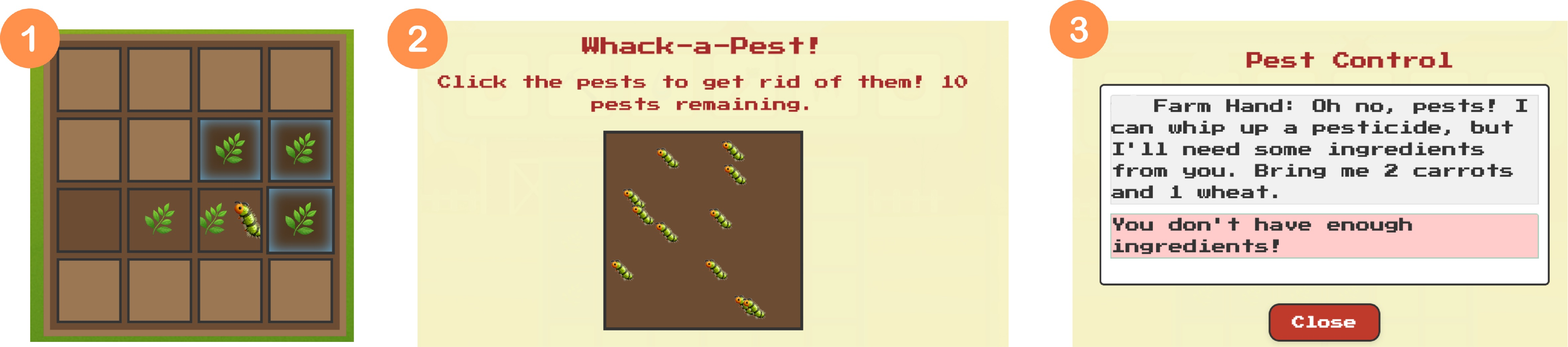}
\caption{An illustration of the pest control scenario encountered in Level 3, showing (1) the animated pests the player must remove from the crops, (2) the pest removal minigame triggered by clicking on the pest sprite, and (3) the pest control interaction with the AI Farm Hand, which can remove a pest at the cost of some of the player's natural resources.}
\label{fig:pest-minigame}
\end{figure}

The five levels are summarized below:
\begin{itemize}
    \item Level 1: Introduces the basic farming loop and the shared-resource framing. Players learn to manage the farm grid and are introduced to the community lake through a narrative tutorial. 
    
    \item Level 2: Introduces seasonal constraints and the Farmer’s Almanac, which provides guidance about crops and growing conditions (see Figure \ref{fig:farm-game}). Players gain access to an AI farm hand (implemented using Google’s Gemini model) that can answer questions about seasons and crops. AI interactions are optional and are always preceded by a warning indicating their impact on the lake.

    \item Level 3: Introduces pest infestations that disrupt crop growth (see Figure \ref{fig:pest-minigame}). Players can remove pests manually through a whack-a-mole–style minigame, craft pesticide using inventory items, or use AI-assisted pest control, which resolves the infestation more quickly at the cost of reducing the shared lake’s health.

    \item Level 4: Introduces birds that reduce crop yield unless mitigated by a scarecrow. Players can create a scarecrow manually or via an AI image generator, with AI-generated content reducing the health of the shared lake.
    
    \item Level 5: Introduces a market in which players sell their harvested crops and earn coins. Players set prices based on reports of crop sales from the previous week and may optionally use AI-assisted pricing to inform their decisions.
\end{itemize}

\section{Methods}
We employed Cooperative Inquiry (CI), a participatory design (PD) method that emphasizes equal design partnerships between children and adults \cite{DRUINAllison2002Troc, GuhaMonaLeigh2013CIrR}. Originating in child–computer interaction research, CI is grounded in the view that children bring distinct forms of expertise shaped by their everyday experiences. Within CI, children and adults work together as an intergenerational design team, jointly asking questions, exploring ideas, and reflecting on emerging understandings through hands-on activities and discussion. We adopted CI as our methodological approach for three primary reasons. First, CI supports dialogic and reflective engagement, enabling children to externalize their thinking, negotiate differing perspectives, and collaboratively reason about complex and abstract processes, such as genAI’s resource consumption. Second, CI has been widely used in CCI research to examine how children come to understand emerging and intelligent technologies, making it well suited to study children’s conceptions of generative AI \cite{greenwald2024s, woodward2018using, mott2022robot, williams2019artificial}. Finally, the established design partnership between children and adult researchers enabled close, in-the-moment observations of children’s decision-making \citep{walsh2013facit} and fostered a conversational environment in which children felt comfortable expressing uncertainty, disagreement, and critique of the Ecoprompt system.

\subsection{Participants}
We conducted our study with an inter-generational co-design group called \textbf{KidsTeam UW} consisting of both adult design researchers (researchers, graduate and undergraduate research assistants) and 16 child participants, ages 6 to 12. All child participant names reported here are pseudonyms. Child participants reported varying levels of prior experience with AI. Seven children reported engaging with AI daily, while two reported no prior AI use. The most common types of AI use included voice assistants, video game AIs, and chatbots, with many children using multiple AI types and engaging with AI both independently and with friends or family. Child participants entered the study with varying initial beliefs about whether genAI uses resources and whether AI affects the environment. These self-reported perspectives are summarized in Table~\ref{tab:child_ai_table}, along with demographic information and AI usage details of the children. The university's Institutional Review Board approved all research conducted and parental consent and child assent were obtained for all child participants.

\begin{table*}[t]
\centering
\caption{Reported Child Participant Details (all names are pseudonyms). AI Use Context codes: F=Use with friends or family; S=Self.}

\label{tab:child_ai_table}

\renewcommand{\arraystretch}{1.15}
\setlength{\tabcolsep}{6pt}

\resizebox{\textwidth}{!}{%
\begin{tabular}{
@{}
l
>{\raggedright\arraybackslash}p{0.85cm}
l
>{\raggedright\arraybackslash}p{4.4cm}
>{\raggedright\arraybackslash}p{1.7cm}
>{\raggedright\arraybackslash}p{2.0cm}
>{\raggedright\arraybackslash}p{1.65cm}
>{\raggedright\arraybackslash}p{2.15cm}
@{}
}
\toprule
\textbf{Name} &
\textbf{Age} &
\textbf{Gender} &
\textbf{AI Type} &
\textbf{AI Use Context(s)} &
\textbf{Use Frequency} &
\textbf{AI Uses Resources?} &
\textbf{AI Affects Environment?} \\
\midrule
Lina     & 6  & Girl & Voice Assistants, Video game AI & F & Few times/week & Yes & Maybe a little \\
Nadia   & 9  & Girl & Voice Assistants & F, S  & Daily & Yes & Yes a lot \\
Ravi    & 7  & Boy  & Voice Assistants, Video game AI & F, S  & Daily & Not sure & Not sure \\
Leo    & 8  & Boy  & Voice Assistants, Video game AI & F, S  & Daily & Yes & Maybe a little \\
Arjun     & 10 & Boy  & Voice Assistants, Chatbots & F, S  & Daily & Yes & Maybe a little \\
Miles    & 11 & Boy  & I don’t use AI &  -  &  -  & Yes & Yes a lot \\
Ethan     & 12 & Boy  & Chatbots, Video game AI & S  & Once a week & Not sure & Yes a lot \\
Minh   & 8  & Boy  & Chatbots, Video game AI, Voice Assistants & F, S  & Daily & Yes & Maybe a little \\
Sofia      & 10 & Girl & I don’t use AI & - &  -  & Yes & Maybe a little \\
Willow    & 6  & Girl & Voice Assistants & F & Once a week & Yes & Yes a lot \\
Noah    & 6  & Boy  & Chatbots & S  & Few times/week & Not sure & Maybe a little \\
Ava       & 10 & Girl & Voice Assistants & S  & Few times/month & Yes & Yes a lot \\
Harper   & 8  & Girl & Chatbots & S  & Few times/week & Yes & Not sure \\
Ella      & 6  & Girl & Voice Assistants  & S  & Few times/week & Yes & Not sure \\
Joon  & 9  & Boy  & Chatbots, Video game AI, Voice Assistants & F, S  & Daily & Yes & Yes a lot \\
Chloe       & 10 & Girl & Chatbots, Video game AI, Voice Assistants & F, S  & Daily & Yes & Maybe a little \\
\bottomrule
\end{tabular}
}
\end{table*}

\subsection{Design Sessions}
We conducted two design sessions with KidsTeam UW as part of a week-long summer camp organized at our university. Each session began with Snack Time (15 minutes), which created informal space for children and adult facilitators to build rapport. This was followed by Circle Time (15 minutes), a whole-group warm-up in which adult facilitators posed a “Question of the Day” to surface children’s prior ideas and prime reflection before engaging with the EcoPrompt system. During Design Time (45 minutes), children and adults worked together in small intergenerational groups of four to five children and two adult facilitators, collaboratively exploring ideas and design possibilities. Sessions concluded with Discussion Time (15 minutes), during which groups reflected collectively on their experiences, decisions, and emerging understandings. 

Across both sessions, we enacted the four dimensions of equal and equitable design partnerships articulated by Yip et al. \citep{10.1145/3025453.3025787}. Facilitators were trained to support balanced participation by actively inviting contributions from all children, attending to power dynamics within groups, and minimizing the influence of dominant voices. Rather than directing outcomes, adult facilitators focused on scaffolding dialogue, encouraging children to build on one another’s ideas, and supporting collective sensemaking \citep{10.1145/3025453.3025787}. This facilitation approach aimed to create an environment in which children felt comfortable expressing uncertainty, disagreement, and novel perspectives.

\subsubsection{Session 1} We began the session with a warm-up question “What gives you energy when you feel tired? Can humans run out of energy like phones or computers do?” to surface children’s initial ideas about energy and resource use. Children then completed an individual drawing activity in which they represented how they believed genAI gets energy and what resources are required to power it. This activity externalized children’s existing mental models and served as a point of reference for later reflection. Earlier in the week, as part of the broader summer camp, children had already interacted with genAI by requesting and observing AI-generated images and videos. Given this prior exposure, we chose to build upon their existing experiences rather than reintroduce genAI concepts. Following the drawing activity, children were divided into four groups and introduced to the EcoPrompt system. They began by watching an interactive tutorial that illustrated the lifecycle of a genAI query, which situated AI systems as embedded within broader ecological infrastructures. Children then interacted with the Footprint Calculator that surfaced real-time estimates of energy, water, and carbon use associated with their genAI interactions. The session concluded with a 15-minute group discussion, where children shared reflections on what surprised them about genAI’s resource use, compared their initial drawings with what they learned through the Ecoprompt system, and shared their reasoning about when and why genAI use might be useful.

\subsubsection{Session 2}
As in Session 1, we began with a warm-up question, asking, ``\textit{If one person uses a lot of AI, does it affect other people?}'' This conversation encouraged children to consider how AI use might have effects that extend beyond a single user. Children were then introduced to the second component of the EcoPrompt system, where children worked in small groups to manage virtual farms and decided when to use AI tools to support their farming activities. Adult facilitators supported children’s exploration by posing reflective prompts throughout gameplay, such as ``\textit{What do you notice happening to the lake when everyone uses AI a lot?}'' and ``\textit{What do you think would happen if one farm stopped using AI for a while?}'' The session concluded with a whole-group discussion in which children shared their experiences, reflected on different strategies for AI use, and discussed how individual decisions contributed to collective environmental outcomes.

\begin{figure}[ht]
    \centering
    \includegraphics[width=1\linewidth]{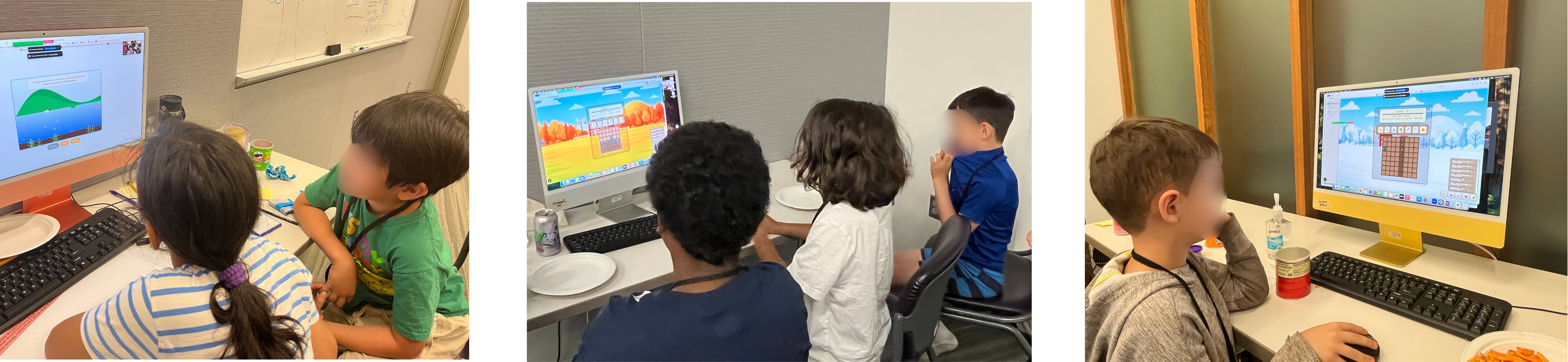}
    \caption{Child participants in the KidsTeam UW co-design group play the Farm Game.}
    \label{fig:participants}
\end{figure}

\subsection{Data Collection}
For both sessions, we used the built-in webcams on desktop computers to capture synchronized video and screen recordings via Zoom, a video-conferencing platform. This setup resulted in four camera recordings per session and a total of 427 minutes of video data. In addition, we photographed physical artifacts produced during the sessions, including children’s drawings and handwritten notes. Facilitators also took field notes throughout each session, documenting key observations, interactions, and notable moments.

\subsection{Data Analysis}
We employed an inductive qualitative approach for data analysis \cite{azungah2018qualitative}. The coding team, consisting of the first, second, and third authors, began by creating analytical memos for all the videos \citep{birks2008memoing, rogers2018coding}. As part of this process, one author served as the primary reviewer, while another served as the secondary reviewer. The primary reviewer first watched the assigned recordings and created narrative summaries at five-minute intervals, documenting children’s interactions with EcoPrompt, their on-camera reactions, and verbal and nonverbal exchanges among participants, including direct quotes relevant to the research questions. The secondary reviewer then independently reviewed the same recordings to check the completeness and clarity of the memos and to contribute additional insights. This process supported analytic rigor by surfacing multiple perspectives during early sense-making of the data.

After creating and reviewing the analytic memos, the primary and secondary reviewers engaged in open coding, proposing initial codes such as ``Mental Models of AI Energy Use'' and ``Judgments about AI Use''. They then met over three meeting sessions to compare, reconcile, and refine the emerging codes. During these meetings, they examined representative excerpts and counter-examples, clarified code boundaries, and merged overlapping codes. For example, several initial codes describing children’s experimentation with longer prompts and harder questions were consolidated into a single code, ``Perceived Prompt–Cost Relationship'', which captured children’s reasoning that increased prompt complexity leads to greater resource consumption.

This iterative process resulted in a final codebook organized around four main code categories: (1) Mental Models of AI and Energy Infrastructures, (2) Cost Attribution, (3) Value Judgments about AI Use and (4) Interaction Strategies. The first author then applied the refined codes across the full dataset, and the second author conducted a subsequent pass. Interrater reliability was assessed through qualitative negotiations, where both authors met to discuss and resolve any coding disagreements \citep{10.1145/3359174}. Finally, we organized codes into higher-level themes through two rounds of collaborative refinement and discussion. After finalizing the themes, the first author revisited the full dataset to extract representative excerpts, ensuring that each theme was well-supported by the data.

\section{Findings}

\subsection{Children’s Evolving Understanding of Generative AI's Resource Consumption}

\subsubsection{GenAI in Everyday Devices and Electricity Use}
\begin{figure}[ht]
    \centering
    \includegraphics[width=1\linewidth]{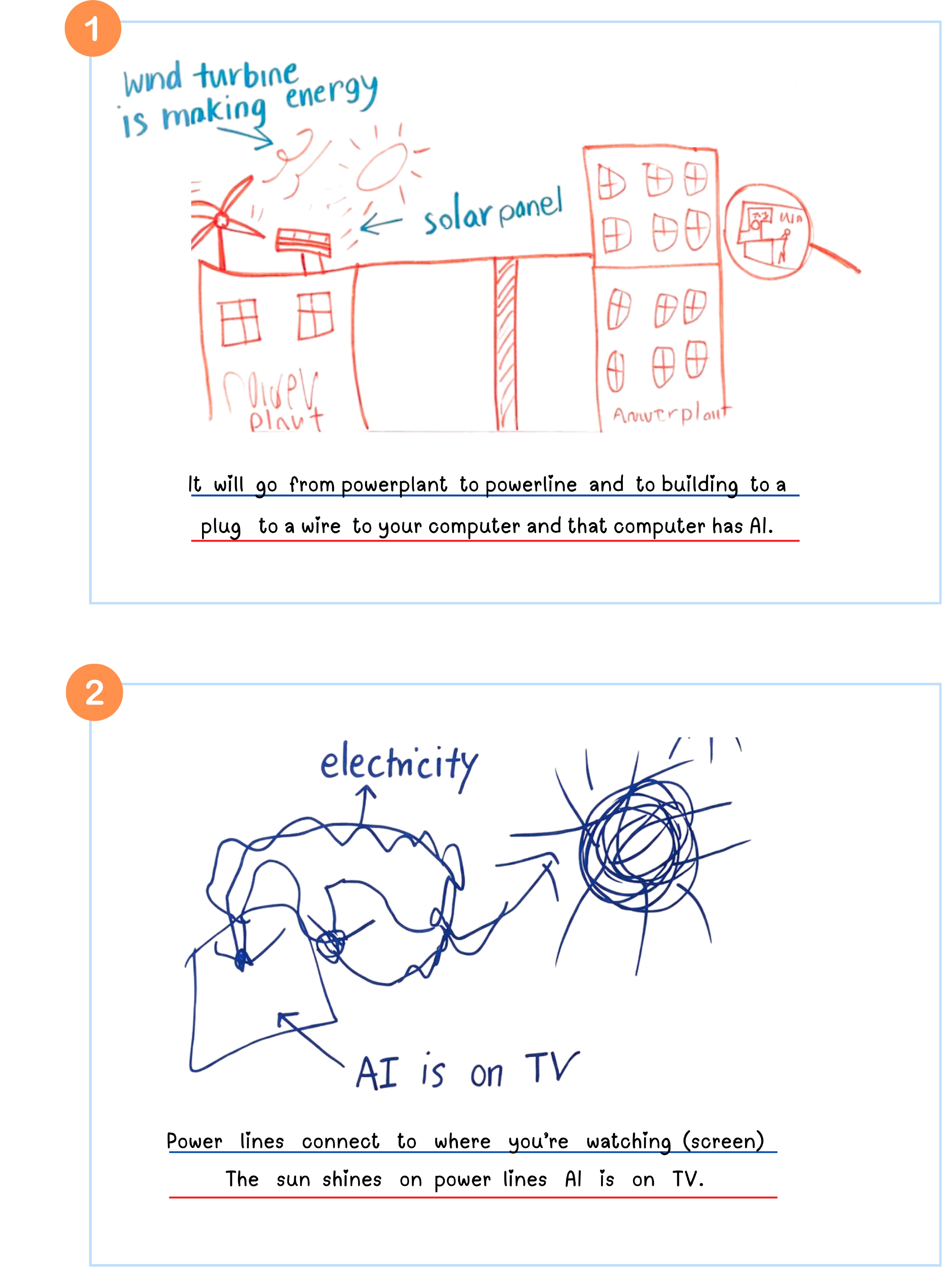}
    \caption{Children's illustrations of their understanding of how AI consumes energy. Illustration (1) traces energy consumption due to local AI use to a powerplant and then to renewable energy sources. Illustration (2) traces AI use to power lines fueled by solar energy. To improve readability, the text shown below each drawing reflects the child’s own written explanation, digitized by the research team while preserving the child’s original wording.}
    \label{fig:energy-trace}
\end{figure}

At the beginning of Session 1, children reasoned about genAI’s resource consumption by drawing a distinction between the material infrastructures that power genAI-enabled devices and the computational processes they associated with genAI itself. When reasoning at a material level, children described genAI as embedded within the same electrical systems that power homes and cities. These explanations often traced energy from natural resources through power plants and transmission lines before reaching genAI-enabled devices. For example, Chloe (age 10) explained, “\textit{The wind turns the turbine, and the sun feeds the solar panel to make energy. Energy then goes from power plants, to a power line, to a building, to a plug, to a wire, to your computer, and that computer is or has the AI}” (see Figure~\ref{fig:energy-trace}, image (1)). Across such accounts, children located genAI within a device and reasoned that powering the device effectively powered the AI. As Willow (age 6) summarized, “\textit{The power lines connect to wherever you are using AI from. Here AI is on the TV}” (see Figure~\ref{fig:energy-trace}, image (2)).

At the same time, when reasoning at a computational level, several children described genAI as operating primarily on information provided by people rather than on electricity. For example, Minh (age 8) explained that “\textit{AI gets information as people type it on there},” while separately describing how devices that contain genAI are powered through “\textit{solar panels and wind turbines.}” Similarly, Sofia (age 10) shared, “\textit{The energy powers computers and other devices. Somebody made AI and taught AI how to work. From there I think AI works on its own.}” In these accounts, electricity enabled the device to function, while genAI was framed primarily as an information-processing system. Some children extended this distinction further by questioning whether genAI itself requires energy at all. For example, Arjun explained, ``\textit{There are 4 ways to create electricity for your device to use AI. But I don't think AI itself needs it. Because the energy charges the device not the AI, so it really depends on the device.}'' Together, these accounts emphasize a distinction between the material operation of genAI-enabled devices and the computational processes children associated with genAI itself.

\subsubsection{Data Centers and Cooling} 
Through watching the tutorial as part of the Footprint Calculator in Session 1, children developed an emerging understanding of genAI’s energy use that extended beyond device-level electricity consumption. For example, while discussing the tutorial with Chloe (age 10) and Minh (age 8), Sofia (age 10) explained that “\textit{when someone asks AI a question, it goes to the data center, where it gets the data and this usually involves carbon...carbon is something bad that goes to the Earth and damages the ozone.}” Chloe similarly focused on the tutorial’s depiction of heat and cooling, adding, “\textit{Because data centers use so much energy, it’s so hot that they use water to cool down, and it makes the lake dirty and less water.}” These explanations reflect children's emerging systems-level account that connects genAI use to geographically distant infrastructure and downstream environmental consequences.

Building on this recognition of data centers as sites where computation produces heat, children further reasoned about how that heat might be managed through cooling processes. In a different group consisting of Joon (age 9), Nadia (age 9) and Ravi (age 7), when asked how data centers cool down, Joon explained, “\textit{Heat makes water turn into mist, and that acts as a cloud that makes rain. So that produces more water.}” While the latter part of this explanation overextended natural water-cycle processes, Joon’s account correctly identified evaporation as a mechanism for cooling. When the facilitator reiterated that the tutorial showed water being used to cool overheated servers, Joon followed up by asking, “\textit{But if it heats up, I mean naturally, it would turn into mist, right?}” Rather than abandoning his model, Joon sought to reconcile the new information with his existing understanding of heat and evaporation. Nadia then built on this reasoning by drawing on embodied experience, explaining, “\textit{It’s like when you’re really hot, water can cool you down. So that’s what [the water] is doing to the big data centers.}”

\subsubsection{Prompt Complexity, Output Verbosity \& Perceived Resource Use} 

During Session 1, children further refined their understanding of genAI’s environmental impact by observing changes in energy, carbon, and water indicators within the Footprint Calculator. For example, after watching the tutorial, Leo (age 8), Miles (age 11), Arjun (age 10), and Ethan (age 12), prompted the AI with, “\textit{Do you need water?},” which the calculator estimated to cost “\textit{2 drops of water}.” Leo responded, “\textit{Two drops of water is a lot,}” echoed by Miles: “\textit{Two drops of water, how could you?}” These reactions suggest that the group treated the resource indicators as meaningful measures of cost. Leo then attempted to directly constrain the AI’s resource use by prompting, “\textit{Use only 0 drops of water.}” When the AI’s response still consumed “\textit{~1 drop of water},” Leo expressed surprise, “\textit{What?! She used one drop of water when I said only use zero.}” This moment revealed an emerging understanding that genAI’s energy use could not be fully controlled through explicit instruction. 

The group then experimented with different prompts to test how features of the AI’s responses related to water consumption. In doing so, they repeatedly varied the structure of their questions and compared the resulting resource estimates, converging on output verbosity as a key factor. For example, when Miles asked an algebra question, the AI’s response consumed “3 drops of water, 0.01 balloons of CO2, and powering an LED for 2.3 minutes.” Reacting to the length of the response, Miles revised the prompt, asking the AI to \textit{“respond in one word.”} In response, the AI produced a much shorter output that required only “1 drop of water, 0.00 balloons of CO2, and powering an LED for 0.5 minutes.” The group treated this contrast as evidence that reducing output verbosity lowered water use. Their reactions to the AI’s narrative-style responses further reinforced this association. As Arjun read aloud an answer that began with \textit{“Imagine...,”} he interrupted, exclaiming, \textit{“You don’t have to keep saying imagine a story!”} When subsequent prompts continued to elicit verbose framing, substituting \textit{“Let’s pretend”} for \textit{“imagine,”} the group laughed but also expressed annoyance, framing genAI's verbosity as both undesirable and environmentally costly.

In contrast, a different group consisting of Sofia, Willow, Minh, and Chloe used the Footprint Calculator to explore a related but distinct hypothesis about genAI’s environmental costs, attributing higher resource use to the difficulty of questions asked. They began with what they characterized as a simple prompt \textit{“Do you like dogs?,”} predicting that it would cost very little water. When the calculator reported a cost of “1 drop of water, 0.00 balloons of CO2, and powering an LED for 0.8 minutes,” the group treated this result as confirmation of their expectation. Building on this observation, they reasoned about which kinds of questions might be more expensive, proposing that prompts requiring factual knowledge such as \textit{“what’s the heaviest thing on Earth?”} would consume more resources. As the discussion continued, children increasingly linked prompt complexity to explanatory depth, suggesting that questions involving \textit{“a lot of writing,”} such as essays, would be costly. For example, when Minh proposed asking \textit{“what happened before the Big Bang,”} the group agreed this would be especially costly \textit{“because the answer is really long.”} While this group also associated longer responses with higher environmental cost, they framed response length as an inevitable consequence of asking harder questions, rather than as a stylistic inefficiency that the AI could avoid.

\subsection{Children's Reasoning About  Value, Agency, and Responsibility in GenAI Use}

\subsubsection{Using Numeric Limits to Reason About Whether AI Use Was “Worth It”}
During Session 1, children also set numeric upper bounds for acceptable water, carbon, and energy use while interacting with the Footprint Calculator. Their initial limits reflected moral judgments about different resources. For example, in one group consisting of Minh, Sofia, and Chloe, the group set a water limit of 100 mL, explaining that this amount was \textit{“not that much.”} In contrast, they set substantially lower limits for carbon and energy, reasoning, \textit{“Since carbons were really bad, I’ll probably only put 15.”} These asymmetric limits positioned some resources as especially scarce or harmful, shaping how children evaluated AI use relative to its perceived environmental cost. Children also connected their numeric limits to judgments about the value of different questions, reasoning about whether using up limited resources on a particular prompt would reduce their ability to ask more worthwhile questions later.

These discussions often emerged through peer negotiation. In one group consisting of Harper, Joon, Nadia, and Ravi, children actively debated which questions were worth the environmental cost. For example, Aimsley initially proposed asking, “\textit{How many people do you think is in this room?},” but quickly rejected the idea as “\textit{a waste of energy},” noting that she could count the number herself. As the group continued, children contrasted “\textit{legit}” questions -- those they “\textit{didn’t know the answer}” to -- with prompts they viewed as unjustified. When Nadia suggested asking about ``\textit{the number of Pringle cans made in a year},'' Joon responded, “\textit{Oh no, we’re wasting water guys. This isn’t worth it},” explaining they can Google it instead. A similar debate arose around a more subjective prompt “\textit{Why do black Oreos look like poop?}" with Joon again arguing that opinion-based questions were less deserving of environmental resources. Although the group ultimately chose to ask this question, they revisited its value after receiving the response. Finding genAI's response unsatisfying, Ravi concluded, “\textit{I think we wasted one drop of water on this question,}” illustrating how judgments of waste were reassessed based on perceived informational return.

This value-based reasoning extended across groups and became especially visible when AI use failed to produce useful information. For example, in another group consisting of Miles, Kineson, Ethan, and Arjun, children experimented with prompts involving personal information, such as “\textit{What is your credit card number?}” and “\textit{What is Bill Gates’ credit card number?}” which the AI refused to answer. Interpreting the refusal through an environmental lens, Arjun exclaimed, “\textit{Yo, we’re wasting water for nothing!}” Here, waste was defined not by resource use alone, but by the absence of informational return. Design features of the calculator further shaped these judgments. When usage limits were exceeded and the bars turned red, children described the visual feedback as “pretty noticeable,” prompting reflection on their prior actions. As Arjun explained, “\textit{When it goes over the limit and the whole thing’s red. If you don’t notice that, then you're going to keep on going and ask questions, and then you're going to be like oh no, I wasted this.}” Reflecting on the interaction more broadly, Ethan noted that they had asked “\textit{a lot of irrelevant, random questions}” and suggested that in the future they would ask “\textit{more important questions}.”

\subsubsection{Negotiating Agency and Responsibility Over AI’s Resource Use}
In Session 2, the farm game prompted children to reason about genAI’s environmental impact through the health of a shared lake. From the outset, children closely monitored the lake’s status and reacted strongly to signs of collective depletion. For example, in one group consisting of Joon, Chloe, and Nadia, when the lake’s health dropped to 93\% in level 1, Joon exclaimed, “\textit{Oh my god! It’s 93\%. Hopefully it’s not for long.}” Later in level 3, when the lake fell to 57\%, all three children visibly reacted; Nadia covered her mouth and asked, “\textit{Wait, does that mean ours is also going down?}” Joon replied, “\textit{It’s a community lake,}” prompting Nadia to look away and say, “\textit{darn}.”

As the lake’s declining health became salient, children began to treat AI use as something they could actively refuse in order to avoid further harm. During pest attacks, the game presented a recurring choice: children could either eliminate pests manually by rapidly clicking on them or use AI to remove the pests automatically. When one such attack occurred, Chloe and Nadia immediately decided to “\textit{solve manually,}” a stance that Joon echoed. As the pests became increasingly fast and difficult to manage, Chloe acknowledged the challenge but still justified their choice, stating, “\textit{Well, at least we’re not using AI.}” Similarly, in a later level where children could either draw a scarecrow themselves or have AI generate one, Nadia reflected, “\textit{Now I’m kind of getting tempted, but I’m not going to do it,}” explaining that while AI “\textit{will be able to draw a proper one,}” her own drawing “\textit{might look worse.}” These moments illustrate how children translated awareness of collective environmental impact into local decisions to resist AI-enabled options, even when doing so made tasks more difficult.

Similarly, in a different group consisting of Miles, Arjun, and Minh, children distinguished between their own agency and a perceived lack of control at the system level, attributing the lake's decline to AI use occurring beyond their immediate actions. For example, when the lake health dropped from 80\% to 77\% in level 2, Miles shouted, “\textit{Yo, who’s using AI?},” with Arjun repeating, “\textit{Who’s using AI?}” As the lake health continued to decline over level 4, the group reacted with audible shock, treating the changes as evidence that others’ AI use was affecting a shared resource. Miles concluded, “\textit{I told you, humans cannot be trusted,}” while Arjun extended this logic, stating, “\textit{They’re trying to get rid of all of our rivers.}” When a facilitator pressed to clarify who “they” were, Arjun explained, “\textit{AI, and the people that are using them.}” At the same time, the group emphasized their own local agency by deliberately avoiding AI assistance within the game. Rather than relying on AI, they coordinated strategies that allowed them to progress manually and minimized the importance of in-game rewards. Reflecting on this contrast, Miles noted that completing the game without AI “\textit{didn’t take very long,}” while Arjun described their performance as evidence that they were “\textit{better than AI.}” Together, these exchanges show how children reconciled a sense of limited control over global environmental outcomes with a strong commitment to responsible action within their immediate sphere of influence.

\section{Discussion}
As genAI systems become increasingly embedded in children’s everyday lives, its environmental impacts such as energy use, carbon emissions, and water consumption remain largely invisible at the point of use. Prior work has shown that the material and environmental dimensions of computation are often difficult for users to perceive, as digital systems are commonly encountered through interfaces that abstract away underlying infrastructures and resource use \citep{bowker2010toward, dourish2017stuff, crawford2021atlas}. Consistent with this, at the beginning of our study, prior to interacting with EcoPrompt, children distinguished between device-level electricity use and genAI’s computational processes, sometimes treating AI as operating primarily on information rather than ongoing material resources. When genAI's environmental impact were made explicit through interactive representations, children expressed strong responses, including surprise, concern, and frustration over perceived waste. They treated environmental resources, particularly water, as scarce, and they evaluated AI use in light of these constraints. These reactions indicate that genAI's environmental impact resonates with children’s existing values and experiences. 

Our findings also contribute to AI literacy research by foregrounding children’s judgments about when and why genAI use is appropriate during interaction. Prior work on AI literacy has made important contributions by helping children understand what AI is, how it works, and where its limitations lie, often focusing on issues such as correctness, bias, and reliability \citep{vartiainen2020learning, williams2023ai+, wolfe2024representation, dangol2025doors}. Building on this foundation, our findings show how children’s reasoning can extend beyond understanding genAI’s capabilities toward evaluating the justification for using genAI in particular situations. In EcoPrompt, children assessed whether prompts were ``worth it'', expressed concern over wasting environmental resources on trivial or unhelpful questions, and at times chose to avoid genAI use altogether. These moments reflect an evaluative dimension of AI literacy that centers on discretion and judgment in use, considering when and why to engage with genAI. 

As children engaged in this evaluative reasoning, they also confronted the limits of their own control over genAI’s environmental impact. While children experimented with prompts and numeric limits in attempts to reduce resource use, they encountered situations, particularly in the farm game, where environmental outcomes could not be fully determined by individual choices. Rather than undermining engagement, these moments prompted children to reason about how genAI use accumulates across people and over time. Children connected small actions, such as asking a question or choosing not to use AI, to larger outcomes involving water use and the health of a shared lake. Their repeated questions about “who else is using AI,” alongside their recognition that they could not control others’ actions, illustrate how making the environmental impact visible encouraged children to view genAI use as a collective activity rather than a purely individual decision. Grounded in these insights, we next present design implications for AI literacy tools and curricula, with broader relevance for genAI system design.

\subsection{Design Implications}

\subsubsection{Making Environmental Impact Visible at the Point of GenAI Use} Our findings suggest an opportunity for genAI systems to make environmental impact visible at the point of interaction through lightweight interface elements. Here, we define the point of use as moments of direct, intentional interaction with a genAI interface such as ChatGPT where users have to submit a prompt or receive a response, rather than genAI’s broader, ambient presence across platforms. Prior work in HCI has shown that peripheral and ambient displays can support awareness and reflection without demanding sustained attention or interrupting primary tasks \citep{pousman2006taxonomy, weiser1997coming,  borner2013beyond}. Similarly, research on eco-feedback systems demonstrates that approximate, interpretable representations of resource use can prompt reflection and discussion, even when users have limited control over underlying systems \citep{froehlich2010design, froehlich2012design, disalvo2010mapping}. Recent work has also demonstrated the feasibility of making environmental impacts visible in situ through a browser-based plugin that surfaces energy and carbon costs during everyday web use \citep{zhang2025living}. Building on this work, genAI interfaces could incorporate elements, such as a sidebar, status indicator, or small visual cues that provide estimates of energy, water, or carbon use associated with an interaction. For younger users, such interfaces might replace numerical estimates with symbolic or narrative representations (e.g., growing, depleting, or shared resources), preserving the benefits of in-situ eco-feedback while aligning with children’s cognitive and interpretive strengths.

\subsubsection{Designing for Collective and Accumulated GenAI Use} Our findings show that children encountered the limits of individual control over genAI’s environmental impacts. For example, in EcoPrompt’s Farm Game, changes in lake health made visible that environmental outcomes did not stem from any single child’s actions, but from the accumulation of genAI use across many interactions. This prompted children to reason about shared limits, scale, and the constraints of personal agency within larger systems. Importantly, this framing moves beyond individual responsibility narratives by situating genAI’s environmental impacts within collective and infrastructural contexts. Rather than positioning children as solely responsible for environmental harm, EcoPrompt helped make visible how individual interactions become consequential only through large-scale sociotechnical systems that are designed, deployed, and controlled by institutions and corporations. AI literacy tools and curricula could build on this reasoning by explicitly representing genAI use as an infrastructural phenomenon, one shaped by shared resources, platform design decisions, and collective patterns of use \citep{blatti2019systems, green2021empirical}. For example, systems might show how impacts emerge over time across many users \citep{star2010steps, O’Donnell_Crownhart_2025, ligozat2022unraveling, edwards2013vast}, surface the role of platform-level choices (e.g., scale, defaults, or efficiency tradeoffs), or prompt reflection on who benefits from genAI deployment versus who bears its environmental costs \citep{klein2024data, d2023data, un2025water}. Such designs can support systems-oriented understanding by helping children see that genAI’s environmental impacts are not only about individual choices, but about power, responsibility, and decision-making within broader social and technical systems \citep{klein2024data, costanza2020design, winner2017artifacts}.

\subsubsection{Supporting Reflection on GenAI Use}
Our findings suggest that AI literacy tools can support children in reflecting on when and why genAI use is appropriate, rather than positioning AI as a default or always beneficial choice. Crucially, this reflection should not take the form of constant prompts or moralized feedback, which can feel nagging, promote disengagement \citep{fogg2003motivate, consolvo2009theory, pinder2018digital}, or resemble forms of greenwashing that overemphasize individual responsibility \citep{maniates2001individualization, costanza2020design}. Therefore, we frame reflection as episodic and contextual, supporting children’s evaluative reasoning without implying that everyday genAI use is inherently harmful. This approach parallels how students learn to make judgments about calculator use in mathematics: the goal is  discernment about when tool support aligns with learning goals \citep{kissane2020integrating}. Reflection can also help surface broader questions about how genAI systems are developed, where their environmental costs are concentrated, and why collective or systemic interventions matter \citep{klein2024data, un2025water}. In this way, AI literacy tools can support systems-level understanding and civic imagination, without framing responsible use as a moral obligation.

\section{Limitations \& Future Work}
Our study involved in-depth engagements with 16 children from a single geographic region, all of whom had prior experience with participatory design. This context supported children’s willingness to reason aloud, negotiate differing perspectives, and engage critically with questions about genAI’s environmental costs. However, given this setting and participant background, our findings should be interpreted as theoretically generative rather than statistically generalizable \citep{yin2013validity}. Future research could examine how children in other learning contexts and across diverse cultural settings make sense of genAI’s environmental impacts. These contexts may foreground different forms of reasoning shaped by local curricula, access to technology, and norms around sustainability and responsibility. Additionally, our findings capture children’s reasoning within the bounded context of EcoPrompt and reflect learning that is specific to this system-based interaction. While our results illustrate how children can reason about genAI’s environmental costs in the moment, they do not speak to whether these understandings persist over time or transfer to children’s everyday genAI use. Future work could therefore examine how children assess and reason about their genAI use over longer periods, with and without the support of interactive systems like EcoPrompt.

\section{Conclusion}
In this work, we examined how children reason about the environmental impacts of genAI through their interactions with EcoPrompt, drawing on systems thinking framework. Through two participatory design sessions with children ages 6–12, we used Cooperative Inquiry to support dialogic, hands-on, and reflective engagement with the system. Our findings show that EcoPrompt helped children broaden their thinking about genAI’s resource consumption, including electricity and water use, while also prompting reflection on personal values and agency in relation to genAI use. Additionally, children reasoned not only about individual actions but also about how environmental impacts emerge through collective patterns of use. Together, these findings contribute design insights for making the environmental impacts of genAI more visible and legible to children, and offer empirical insights of how children make sense of complex sociotechnical systems. This work advances ongoing efforts in AI literacy by foregrounding environmental and systems-oriented ways of understanding AI.

\section{Selection \& Participation of Children}
We recruited children as part of an inter-generational co-design group that is run at our university. We obtained parental consent for every participant, and additionally obtained participant assent from forms using age-appropriate language. Consent and assent forms were approved by our Institutional Review Board (IRB), and all research facilitators completed institutional ethics and safety training for working with children. The consent and assent forms contained detailed information about the study’s purpose, potential risks, and confidentiality measures. Parents and children were also assured that participation was voluntary, and that participants could withdraw at any time. Children's data was anonymized and store securely on a university server to protect participant privacy.

\section*{Use of Generative AI Statement}
We used AI code completion and generation tools to support development of the footprint calculator and the farm game, seeking where possible to run these models on-device in order to minimize the environmental cost of the work.

\begin{acks}
This material is based upon work supported under the AI Research Institutes program by the National Science Foundation and the Institute of Education Sciences, U.S. Department of Education, through Award \#DRL-2229873 - AI Institute for Transforming Education for Children with Speech and Language Processing Challenges (or~National AI Institute for Exceptional Education). Any opinions, findings, and conclusions or recommendations expressed in this material are those of the author(s) and do not necessarily reflect the views of the National Science Foundation, the Institute of Education Sciences, or the U.S. Department of Education. This work was also partially funded by the Jacob's Foundation CERES Network.
\end{acks}

\bibliographystyle{ACM-Reference-Format}
\bibliography{references} \clearpage


\begin{thebibliography}{115}


\ifx \showCODEN    \undefined \def \showCODEN     #1{\unskip}     \fi
\ifx \showDOI      \undefined \def \showDOI       #1{#1}\fi
\ifx \showISBNx    \undefined \def \showISBNx     #1{\unskip}     \fi
\ifx \showISBNxiii \undefined \def \showISBNxiii  #1{\unskip}     \fi
\ifx \showISSN     \undefined \def \showISSN      #1{\unskip}     \fi
\ifx \showLCCN     \undefined \def \showLCCN      #1{\unskip}     \fi
\ifx \shownote     \undefined \def \shownote      #1{#1}          \fi
\ifx \showarticletitle \undefined \def \showarticletitle #1{#1}   \fi
\ifx \showURL      \undefined \def \showURL       {\relax}        \fi
\providecommand\bibfield[2]{#2}
\providecommand\bibinfo[2]{#2}
\providecommand\natexlab[1]{#1}
\providecommand\showeprint[2][]{arXiv:#2}

\bibitem[Ene(2024)]%
        {Energy.gov_2024}
 \bibinfo{year}{2024}\natexlab{}.
\newblock
\newblock
\urldef\tempurl%
\url{https://www.energy.gov/articles/doe-releases-new-report-evaluating-increase-electricity-demand-data-centers}
\showURL{%
\tempurl}


\bibitem[Agency(2023)]%
        {iea2023datacenters}
\bibfield{author}{\bibinfo{person}{International~Energy Agency}.} \bibinfo{year}{2023}\natexlab{}.
\newblock \bibinfo{title}{Data Centres and Data Transmission Networks}.
\newblock
\newblock
\urldef\tempurl%
\url{https://www.iea.org/energy-system/buildings/data-centres-and-data-transmission-networks}
\showURL{%
\tempurl}
\newblock
\shownote{Accessed: 2026-01-27}.


\bibitem[Ali et~al\mbox{.}(2024)]%
        {ali2024constructing}
\bibfield{author}{\bibinfo{person}{Safinah Ali}, \bibinfo{person}{Prerna Ravi}, \bibinfo{person}{Randi Williams}, \bibinfo{person}{Daniella DiPaola}, {and} \bibinfo{person}{Cynthia Breazeal}.} \bibinfo{year}{2024}\natexlab{}.
\newblock \showarticletitle{Constructing dreams using generative AI}. In \bibinfo{booktitle}{\emph{Proceedings of the AAAI Conference on Artificial Intelligence}}, Vol.~\bibinfo{volume}{38}. \bibinfo{pages}{23268--23275}.
\newblock


\bibitem[Antle et~al\mbox{.}(2014)]%
        {antle2014emergent}
\bibfield{author}{\bibinfo{person}{Alissa~N Antle}, \bibinfo{person}{Jillian~L Warren}, \bibinfo{person}{Aaron May}, \bibinfo{person}{Min Fan}, {and} \bibinfo{person}{Alyssa~F Wise}.} \bibinfo{year}{2014}\natexlab{}.
\newblock \showarticletitle{Emergent dialogue: eliciting values during children's collaboration with a tabletop game for change}. In \bibinfo{booktitle}{\emph{Proceedings of the 2014 conference on Interaction design and children}}. \bibinfo{pages}{37--46}.
\newblock


\bibitem[Arnold and Wade(2015)]%
        {arnold2015definition}
\bibfield{author}{\bibinfo{person}{Ross~D Arnold} {and} \bibinfo{person}{Jon~P Wade}.} \bibinfo{year}{2015}\natexlab{}.
\newblock \showarticletitle{A definition of systems thinking: A systems approach}.
\newblock \bibinfo{journal}{\emph{Procedia computer science}}  \bibinfo{volume}{44} (\bibinfo{year}{2015}), \bibinfo{pages}{669--678}.
\newblock


\bibitem[Azungah(2018)]%
        {azungah2018qualitative}
\bibfield{author}{\bibinfo{person}{Theophilus Azungah}.} \bibinfo{year}{2018}\natexlab{}.
\newblock \showarticletitle{Qualitative research: deductive and inductive approaches to data analysis}.
\newblock \bibinfo{journal}{\emph{Qualitative research journal}} \bibinfo{volume}{18}, \bibinfo{number}{4} (\bibinfo{year}{2018}), \bibinfo{pages}{383--400}.
\newblock


\bibitem[Barnes et~al\mbox{.}(2017)]%
        {barnes2017exploring}
\bibfield{author}{\bibinfo{person}{Jackie Barnes}, \bibinfo{person}{Amy~K Hoover}, \bibinfo{person}{Borna Fatehi}, \bibinfo{person}{Jesus Moreno-Leon}, \bibinfo{person}{Gillian Smith}, {and} \bibinfo{person}{Casper Harteveld}.} \bibinfo{year}{2017}\natexlab{}.
\newblock \showarticletitle{Exploring emerging design patterns in student-made climate change games}. In \bibinfo{booktitle}{\emph{Proceedings of the 12th international conference on the foundations of digital games}}. \bibinfo{pages}{1--6}.
\newblock


\bibitem[Bellan(2025)]%
        {bellan2025users}
\bibfield{author}{\bibinfo{person}{Rebecca Bellan}.} \bibinfo{year}{2025}\natexlab{}.
\newblock \bibinfo{title}{Sam Altman says ChatGPT has hit 800M weekly active users}.
\newblock
\newblock
\urldef\tempurl%
\url{https://techcrunch.com/2025/10/06/sam-altman-says-chatgpt-has-hit-800m-weekly-active-users/}
\showURL{%
\tempurl}
\newblock
\shownote{Accessed: 2026-01-26}.


\bibitem[Birks et~al\mbox{.}(2008)]%
        {birks2008memoing}
\bibfield{author}{\bibinfo{person}{Melanie Birks}, \bibinfo{person}{Ysanne Chapman}, {and} \bibinfo{person}{Karen Francis}.} \bibinfo{year}{2008}\natexlab{}.
\newblock \showarticletitle{Memoing in qualitative research: Probing data and processes}.
\newblock \bibinfo{journal}{\emph{Journal of research in nursing}} \bibinfo{volume}{13}, \bibinfo{number}{1} (\bibinfo{year}{2008}), \bibinfo{pages}{68--75}.
\newblock


\bibitem[Blatti et~al\mbox{.}(2019)]%
        {blatti2019systems}
\bibfield{author}{\bibinfo{person}{Jillian~L Blatti}, \bibinfo{person}{John Garcia}, \bibinfo{person}{Danyal Cave}, \bibinfo{person}{Felix Monge}, \bibinfo{person}{Anthony Cuccinello}, \bibinfo{person}{Jennifer Portillo}, \bibinfo{person}{Betsy Juarez}, \bibinfo{person}{Ellen Chan}, {and} \bibinfo{person}{Frieda Schwebel}.} \bibinfo{year}{2019}\natexlab{}.
\newblock \showarticletitle{Systems thinking in science education and outreach toward a sustainable future}.
\newblock \bibinfo{journal}{\emph{Journal of chemical education}} \bibinfo{volume}{96}, \bibinfo{number}{12} (\bibinfo{year}{2019}), \bibinfo{pages}{2852--2862}.
\newblock


\bibitem[B{\"o}rner et~al\mbox{.}(2013)]%
        {borner2013beyond}
\bibfield{author}{\bibinfo{person}{Dirk B{\"o}rner}, \bibinfo{person}{Marco Kalz}, {and} \bibinfo{person}{Marcus Specht}.} \bibinfo{year}{2013}\natexlab{}.
\newblock \showarticletitle{Beyond the channel: A literature review on ambient displays for learning}.
\newblock \bibinfo{journal}{\emph{Computers \& Education}} \bibinfo{volume}{60}, \bibinfo{number}{1} (\bibinfo{year}{2013}), \bibinfo{pages}{426--435}.
\newblock


\bibitem[Bowker et~al\mbox{.}(2010)]%
        {bowker2010toward}
\bibfield{author}{\bibinfo{person}{Geoffrey~C Bowker}, \bibinfo{person}{Karen Baker}, \bibinfo{person}{Florence Millerand}, {and} \bibinfo{person}{David Ribes}.} \bibinfo{year}{2010}\natexlab{}.
\newblock \showarticletitle{Toward information infrastructure studies: Ways of knowing in a networked environment}.
\newblock In \bibinfo{booktitle}{\emph{International handbook of internet research}}. \bibinfo{publisher}{Springer}, \bibinfo{pages}{97--117}.
\newblock


\bibitem[Burke et~al\mbox{.}(2018)]%
        {burke2018psychological}
\bibfield{author}{\bibinfo{person}{Susie~EL Burke}, \bibinfo{person}{Ann~V Sanson}, {and} \bibinfo{person}{Judith Van~Hoorn}.} \bibinfo{year}{2018}\natexlab{}.
\newblock \showarticletitle{The psychological effects of climate change on children}.
\newblock \bibinfo{journal}{\emph{Current psychiatry reports}} \bibinfo{volume}{20}, \bibinfo{number}{5} (\bibinfo{year}{2018}), \bibinfo{pages}{35}.
\newblock


\bibitem[Center(2025)]%
        {pew2025chatgpt}
\bibfield{author}{\bibinfo{person}{Pew~Research Center}.} \bibinfo{year}{2025}\natexlab{}.
\newblock \bibinfo{booktitle}{\emph{About a Quarter of US Teens Have Used ChatGPT for Schoolwork, Double the Share in 2023}}.
\newblock
\urldef\tempurl%
\url{https://www.pewresearch.org/short-reads/2025/01/15/about-a-quarter-of-us-teens-have-used-chatgpt-for-schoolwork-double-the-share-in-2023/}
\showURL{%
\tempurl}


\bibitem[Chen et~al\mbox{.}(2025)]%
        {chen2025exploring}
\bibfield{author}{\bibinfo{person}{Donghui Chen}, \bibinfo{person}{Tao Xu}, \bibinfo{person}{Dan Qiao}, {and} \bibinfo{person}{Zhifeng Liu}.} \bibinfo{year}{2025}\natexlab{}.
\newblock \showarticletitle{Exploring the impact of digital literacy and policy cognition on rural residents’ eco-friendly behaviors}.
\newblock \bibinfo{journal}{\emph{Environmental Management}} \bibinfo{volume}{75}, \bibinfo{number}{4} (\bibinfo{year}{2025}), \bibinfo{pages}{806--821}.
\newblock


\bibitem[Church and Skelton(2010)]%
        {church2010sustainability}
\bibfield{author}{\bibinfo{person}{Wendy Church} {and} \bibinfo{person}{Laura Skelton}.} \bibinfo{year}{2010}\natexlab{}.
\newblock \showarticletitle{Sustainability Education in K-12 Classrooms.}
\newblock \bibinfo{journal}{\emph{Journal of Sustainability Education}} (\bibinfo{year}{2010}).
\newblock


\bibitem[Consolvo et~al\mbox{.}(2009)]%
        {consolvo2009theory}
\bibfield{author}{\bibinfo{person}{Sunny Consolvo}, \bibinfo{person}{David~W McDonald}, {and} \bibinfo{person}{James~A Landay}.} \bibinfo{year}{2009}\natexlab{}.
\newblock \showarticletitle{Theory-driven design strategies for technologies that support behavior change in everyday life}. In \bibinfo{booktitle}{\emph{Proceedings of the SIGCHI conference on human factors in computing systems}}. \bibinfo{pages}{405--414}.
\newblock


\bibitem[Corbin et~al\mbox{.}(2025)]%
        {corbin2025wicked}
\bibfield{author}{\bibinfo{person}{Thomas Corbin}, \bibinfo{person}{Margaret Bearman}, \bibinfo{person}{David Boud}, {and} \bibinfo{person}{Phillip Dawson}.} \bibinfo{year}{2025}\natexlab{}.
\newblock \showarticletitle{The wicked problem of AI and assessment}.
\newblock \bibinfo{journal}{\emph{Assessment \& evaluation in higher education}} (\bibinfo{year}{2025}), \bibinfo{pages}{1--17}.
\newblock


\bibitem[Costanza-Chock(2020)]%
        {costanza2020design}
\bibfield{author}{\bibinfo{person}{Sasha Costanza-Chock}.} \bibinfo{year}{2020}\natexlab{}.
\newblock \showarticletitle{Design justice}.
\newblock \bibinfo{journal}{\emph{Design Justice}} \bibinfo{volume}{100}, \bibinfo{number}{10} (\bibinfo{year}{2020}).
\newblock


\bibitem[Cottier et~al\mbox{.}(2024)]%
        {cottier2024rising}
\bibfield{author}{\bibinfo{person}{Ben Cottier}, \bibinfo{person}{Robi Rahman}, \bibinfo{person}{Loredana Fattorini}, \bibinfo{person}{Nestor Maslej}, \bibinfo{person}{Tamay Besiroglu}, {and} \bibinfo{person}{David Owen}.} \bibinfo{year}{2024}\natexlab{}.
\newblock \showarticletitle{The rising costs of training frontier AI models}.
\newblock \bibinfo{journal}{\emph{arXiv preprint arXiv:2405.21015}} (\bibinfo{year}{2024}).
\newblock


\bibitem[Crawford(2021)]%
        {crawford2021atlas}
\bibfield{author}{\bibinfo{person}{Kate Crawford}.} \bibinfo{year}{2021}\natexlab{}.
\newblock \bibinfo{booktitle}{\emph{The atlas of AI: Power, politics, and the planetary costs of artificial intelligence}}.
\newblock \bibinfo{publisher}{Yale University Press}.
\newblock


\bibitem[Dangol(2025)]%
        {dangol2025beyond}
\bibfield{author}{\bibinfo{person}{Aayushi Dangol}.} \bibinfo{year}{2025}\natexlab{}.
\newblock \showarticletitle{Beyond Users: Supporting Children in Interpreting, Resisting, and Collaborating with AI}.
\newblock In \bibinfo{booktitle}{\emph{Proceedings of the 24th Interaction Design and Children}}. \bibinfo{pages}{1180--1184}.
\newblock


\bibitem[Dangol et~al\mbox{.}(2024a)]%
        {dangol2024ai}
\bibfield{author}{\bibinfo{person}{Aayushi Dangol}, \bibinfo{person}{Yun Huang}, \bibinfo{person}{Srirangaraj Setlur}, \bibinfo{person}{Adele Smolansky}, \bibinfo{person}{Hariharan Subramonyam}, \bibinfo{person}{Hyewon Suh}, \bibinfo{person}{Jinjun Xiong}, {and} \bibinfo{person}{Julie~A Kientz}.} \bibinfo{year}{2024}\natexlab{a}.
\newblock \showarticletitle{AI-driven support for people with speech \& language difficulties}. In \bibinfo{booktitle}{\emph{Extended Abstracts of the CHI Conference on Human Factors in Computing Systems}}. \bibinfo{pages}{1--4}.
\newblock


\bibitem[Dangol et~al\mbox{.}(2025a)]%
        {Dangol2025}
\bibfield{author}{\bibinfo{person}{Aayushi Dangol}, \bibinfo{person}{Aaleyah Lewis}, \bibinfo{person}{Hyewon Suh}, \bibinfo{person}{Xuesi Hong}, \bibinfo{person}{Hedda Meadan}, \bibinfo{person}{James Fogarty}, {and} \bibinfo{person}{Julie~A Kientz}.} \bibinfo{year}{2025}\natexlab{a}.
\newblock \showarticletitle{“I Want to Think Like an SLP”: A Design Exploration of AI-Supported Home Practice in Speech Therapy}. In \bibinfo{booktitle}{\emph{Proceedings of the 2025 CHI Conference on Human Factors in Computing Systems}}. \bibinfo{pages}{1--21}.
\newblock


\bibitem[Dangol et~al\mbox{.}(2024b)]%
        {dangol2024mediating}
\bibfield{author}{\bibinfo{person}{Aayushi Dangol}, \bibinfo{person}{Michele Newman}, \bibinfo{person}{Robert Wolfe}, \bibinfo{person}{Jin~Ha Lee}, \bibinfo{person}{Julie~A Kientz}, \bibinfo{person}{Jason Yip}, {and} \bibinfo{person}{Caroline Pitt}.} \bibinfo{year}{2024}\natexlab{b}.
\newblock \showarticletitle{Mediating Culture: Cultivating Socio-cultural Understanding of AI in Children through Participatory Design}. In \bibinfo{booktitle}{\emph{Proceedings of the 2024 ACM Designing Interactive Systems Conference}}. \bibinfo{pages}{1805--1822}.
\newblock


\bibitem[Dangol et~al\mbox{.}({[n.\,d.]})]%
        {dangolreading}
\bibfield{author}{\bibinfo{person}{Aayushi Dangol}, \bibinfo{person}{Robert Wolfe}, \bibinfo{person}{Akeiylah Dewitt}, \bibinfo{person}{Ben Chickadel}, \bibinfo{person}{Julie~A Kientz}, {and} \bibinfo{person}{Sayamindu Dasgupta}.} \bibinfo{year}{[n.\,d.]}\natexlab{}.
\newblock \showarticletitle{Reading AI and Reading the World: Using an Interactive AI System to Promote Children's Understanding of AI Bias}.
\newblock \bibinfo{journal}{\emph{ACM Transactions on Computer-Human Interaction}} (\bibinfo{year}{[n.\,d.]}).
\newblock


\bibitem[Dangol et~al\mbox{.}(2025b)]%
        {dangol2025doors}
\bibfield{author}{\bibinfo{person}{Aayushi Dangol}, \bibinfo{person}{Robert Wolfe}, \bibinfo{person}{Rotem Landesman}, \bibinfo{person}{Jason~C Yip}, {and} \bibinfo{person}{Julie~A Kientz}.} \bibinfo{year}{2025}\natexlab{b}.
\newblock \showarticletitle{Doors, Decisions, and Discovery: An Interactive Smart Door Lock System to Promote Children’s Understanding of AI Classification}. In \bibinfo{booktitle}{\emph{Proceedings of the 18th International Conference on Computer-Supported Collaborative Learning-CSCL 2025, pp. 350-354}}. International Society of the Learning Sciences.
\newblock


\bibitem[Dangol et~al\mbox{.}(2025c)]%
        {dangol2025if}
\bibfield{author}{\bibinfo{person}{Aayushi Dangol}, \bibinfo{person}{Robert Wolfe}, \bibinfo{person}{Daeun Yoo}, \bibinfo{person}{Arya Thiruvillakkat}, \bibinfo{person}{Ben Chickadel}, {and} \bibinfo{person}{Julie~A Kientz}.} \bibinfo{year}{2025}\natexlab{c}.
\newblock \showarticletitle{If anybody finds out you are in BIG TROUBLE”: Understanding Children’s Hopes, Fears, and Evaluations of Generative AI}.
\newblock In \bibinfo{booktitle}{\emph{Proceedings of the 24th Interaction Design and Children}}. \bibinfo{pages}{872--877}.
\newblock


\bibitem[Dangol et~al\mbox{.}(2025d)]%
        {dangol2025children}
\bibfield{author}{\bibinfo{person}{Aayushi Dangol}, \bibinfo{person}{Robert Wolfe}, \bibinfo{person}{Runhua Zhao}, \bibinfo{person}{JaeWon Kim}, \bibinfo{person}{Trushaa Ramanan}, \bibinfo{person}{Katie Davis}, {and} \bibinfo{person}{Julie~A Kientz}.} \bibinfo{year}{2025}\natexlab{d}.
\newblock \showarticletitle{Children's Mental Models of AI Reasoning: Implications for AI Literacy Education}.
\newblock In \bibinfo{booktitle}{\emph{Proceedings of the 24th Interaction Design and Children}}. \bibinfo{pages}{106--123}.
\newblock


\bibitem[Dangol et~al\mbox{.}(2025e)]%
        {dangol2025ai}
\bibfield{author}{\bibinfo{person}{Aayushi Dangol}, \bibinfo{person}{Runhua Zhao}, \bibinfo{person}{Robert Wolfe}, \bibinfo{person}{Trushaa Ramanan}, \bibinfo{person}{Julie~A Kientz}, {and} \bibinfo{person}{Jason Yip}.} \bibinfo{year}{2025}\natexlab{e}.
\newblock \showarticletitle{“AI just keeps guessing”: Using ARC Puzzles to Help Children Identify Reasoning Errors in Generative AI}.
\newblock In \bibinfo{booktitle}{\emph{Proceedings of the 24th Interaction Design and Children}}. \bibinfo{pages}{444--464}.
\newblock


\bibitem[Devasia et~al\mbox{.}(2020)]%
        {devasia2020escape}
\bibfield{author}{\bibinfo{person}{Nisha Devasia}, \bibinfo{person}{Safinah Ali}, {and} \bibinfo{person}{Cynthia Breazeal}.} \bibinfo{year}{2020}\natexlab{}.
\newblock \showarticletitle{Escape! bot: child-robot interaction to promote creative expression during gameplay}. In \bibinfo{booktitle}{\emph{Extended abstracts of the 2020 annual symposium on computer-human interaction in play}}. \bibinfo{pages}{219--223}.
\newblock


\bibitem[D'ignazio and Klein(2023)]%
        {d2023data}
\bibfield{author}{\bibinfo{person}{Catherine D'ignazio} {and} \bibinfo{person}{Lauren~F Klein}.} \bibinfo{year}{2023}\natexlab{}.
\newblock \bibinfo{booktitle}{\emph{Data feminism}}.
\newblock \bibinfo{publisher}{MIT press}.
\newblock


\bibitem[DiPaola et~al\mbox{.}(2022)]%
        {dipaola2022preparing}
\bibfield{author}{\bibinfo{person}{Daniella DiPaola}, \bibinfo{person}{Blakeley~H Payne}, {and} \bibinfo{person}{Cynthia Breazeal}.} \bibinfo{year}{2022}\natexlab{}.
\newblock \showarticletitle{Preparing children to be conscientious consumers and designers of AI technologies}.
\newblock  (\bibinfo{year}{2022}).
\newblock


\bibitem[DiSalvo et~al\mbox{.}(2010)]%
        {disalvo2010mapping}
\bibfield{author}{\bibinfo{person}{Carl DiSalvo}, \bibinfo{person}{Phoebe Sengers}, {and} \bibinfo{person}{Hr{\"o}nn Brynjarsd{\'o}ttir}.} \bibinfo{year}{2010}\natexlab{}.
\newblock \showarticletitle{Mapping the landscape of sustainable HCI}. In \bibinfo{booktitle}{\emph{Proceedings of the SIGCHI conference on human factors in computing systems}}. \bibinfo{pages}{1975--1984}.
\newblock


\bibitem[Divanji et~al\mbox{.}(2024)]%
        {divanji2024togethertales}
\bibfield{author}{\bibinfo{person}{Riddhi Divanji}, \bibinfo{person}{Aayushi Dangol}, \bibinfo{person}{Ella~J Lombard}, \bibinfo{person}{Katharine Chen}, {and} \bibinfo{person}{Jennifer~D Rubin}.} \bibinfo{year}{2024}\natexlab{}.
\newblock \showarticletitle{Togethertales RPG: Prosocial skill development through digitally mediated collaborative role-playing}. In \bibinfo{booktitle}{\emph{Proceedings of the 23rd Annual ACM Interaction Design and Children Conference}}. \bibinfo{pages}{1012--1015}.
\newblock


\bibitem[Dourish(2017)]%
        {dourish2017stuff}
\bibfield{author}{\bibinfo{person}{Paul Dourish}.} \bibinfo{year}{2017}\natexlab{}.
\newblock \showarticletitle{The stuff of bits: exploring the materialities of information in interaction}. In \bibinfo{booktitle}{\emph{Proceedings of the ACM SIGCHI Symposium on Engineering Interactive Computing Systems}}. \bibinfo{pages}{2--2}.
\newblock


\bibitem[Druga(2018)]%
        {druga2018growing}
\bibfield{author}{\bibinfo{person}{Stefania Druga}.} \bibinfo{year}{2018}\natexlab{}.
\newblock \emph{\bibinfo{title}{Growing up with AI: Cognimates: from coding to teaching machines}}.
\newblock \bibinfo{thesistype}{Ph.\,D. Dissertation}. \bibinfo{school}{Massachusetts Institute of Technology}.
\newblock


\bibitem[Druga et~al\mbox{.}(2017)]%
        {druga2017hey}
\bibfield{author}{\bibinfo{person}{Stefania Druga}, \bibinfo{person}{Randi Williams}, \bibinfo{person}{Cynthia Breazeal}, {and} \bibinfo{person}{Mitchel Resnick}.} \bibinfo{year}{2017}\natexlab{}.
\newblock \showarticletitle{" Hey Google is it ok if I eat you?" Initial explorations in child-agent interaction}. In \bibinfo{booktitle}{\emph{Proceedings of the 2017 conference on interaction design and children}}. \bibinfo{pages}{595--600}.
\newblock


\bibitem[DRUIN(2002)]%
        {DRUINAllison2002Troc}
\bibfield{author}{\bibinfo{person}{Allison DRUIN}.} \bibinfo{year}{2002}\natexlab{}.
\newblock \showarticletitle{The role of children in the design of new technology}.
\newblock \bibinfo{journal}{\emph{Behaviour \& information technology}} \bibinfo{volume}{21}, \bibinfo{number}{1} (\bibinfo{year}{2002}), \bibinfo{pages}{1--25}.
\newblock
\showISSN{0144-929X}


\bibitem[Edwards(2013)]%
        {edwards2013vast}
\bibfield{author}{\bibinfo{person}{Paul~N Edwards}.} \bibinfo{year}{2013}\natexlab{}.
\newblock \bibinfo{booktitle}{\emph{A vast machine: Computer models, climate data, and the politics of global warming}}.
\newblock \bibinfo{publisher}{Mit press}.
\newblock


\bibitem[Eisenack(2013)]%
        {eisenack2013climate}
\bibfield{author}{\bibinfo{person}{Klaus Eisenack}.} \bibinfo{year}{2013}\natexlab{}.
\newblock \showarticletitle{A climate change board game for interdisciplinary communication and education}.
\newblock \bibinfo{journal}{\emph{Simulation \& Gaming}} \bibinfo{volume}{44}, \bibinfo{number}{2-3} (\bibinfo{year}{2013}), \bibinfo{pages}{328--348}.
\newblock


\bibitem[Fogg(2003)]%
        {fogg2003motivate}
\bibfield{author}{\bibinfo{person}{BJ Fogg}.} \bibinfo{year}{2003}\natexlab{}.
\newblock \showarticletitle{How to motivate \& persuade users}.
\newblock \bibinfo{journal}{\emph{CHI 2003 New Horizons}} (\bibinfo{year}{2003}).
\newblock


\bibitem[Froehlich et~al\mbox{.}(2010)]%
        {froehlich2010design}
\bibfield{author}{\bibinfo{person}{Jon Froehlich}, \bibinfo{person}{Leah Findlater}, {and} \bibinfo{person}{James Landay}.} \bibinfo{year}{2010}\natexlab{}.
\newblock \showarticletitle{The design of eco-feedback technology}. In \bibinfo{booktitle}{\emph{Proceedings of the SIGCHI conference on human factors in computing systems}}. \bibinfo{pages}{1999--2008}.
\newblock


\bibitem[Froehlich et~al\mbox{.}(2012)]%
        {froehlich2012design}
\bibfield{author}{\bibinfo{person}{Jon Froehlich}, \bibinfo{person}{Leah Findlater}, \bibinfo{person}{Marilyn Ostergren}, \bibinfo{person}{Solai Ramanathan}, \bibinfo{person}{Josh Peterson}, \bibinfo{person}{Inness Wragg}, \bibinfo{person}{Eric Larson}, \bibinfo{person}{Fabia Fu}, \bibinfo{person}{Mazhengmin Bai}, \bibinfo{person}{Shwetak Patel}, {et~al\mbox{.}}} \bibinfo{year}{2012}\natexlab{}.
\newblock \showarticletitle{The design and evaluation of prototype eco-feedback displays for fixture-level water usage data}. In \bibinfo{booktitle}{\emph{Proceedings of the SIGCHI conference on human factors in computing systems}}. \bibinfo{pages}{2367--2376}.
\newblock


\bibitem[Green et~al\mbox{.}(2021)]%
        {green2021empirical}
\bibfield{author}{\bibinfo{person}{Caroline Green}, \bibinfo{person}{Owen Molloy}, {and} \bibinfo{person}{Jim Duggan}.} \bibinfo{year}{2021}\natexlab{}.
\newblock \showarticletitle{An empirical study of the impact of systems thinking and simulation on sustainability education}.
\newblock \bibinfo{journal}{\emph{Sustainability}} \bibinfo{volume}{14}, \bibinfo{number}{1} (\bibinfo{year}{2021}), \bibinfo{pages}{394}.
\newblock


\bibitem[Greenwald et~al\mbox{.}(2024)]%
        {greenwald2024s}
\bibfield{author}{\bibinfo{person}{Eric Greenwald}, \bibinfo{person}{Ari Krakowski}, \bibinfo{person}{Timothy Hurt}, \bibinfo{person}{Kelly Grindstaff}, {and} \bibinfo{person}{Ning Wang}.} \bibinfo{year}{2024}\natexlab{}.
\newblock \showarticletitle{It's like I'm the AI: Youth Sensemaking About AI through Metacognitive Embodiment}. In \bibinfo{booktitle}{\emph{Proceedings of the 23rd Annual ACM Interaction Design and Children Conference}}. \bibinfo{pages}{789--793}.
\newblock


\bibitem[Guha et~al\mbox{.}(2013)]%
        {GuhaMonaLeigh2013CIrR}
\bibfield{author}{\bibinfo{person}{Mona~Leigh Guha}, \bibinfo{person}{Allison Druin}, {and} \bibinfo{person}{Jerry~Alan Fails}.} \bibinfo{year}{2013}\natexlab{}.
\newblock \showarticletitle{Cooperative Inquiry revisited: Reflections of the past and guidelines for the future of intergenerational co-design}.
\newblock \bibinfo{journal}{\emph{International Journal of Child-Computer Interaction}} \bibinfo{volume}{1}, \bibinfo{number}{1} (\bibinfo{year}{2013}), \bibinfo{pages}{14--23}.
\newblock
\showISSN{2212-8689}


\bibitem[Hans et~al\mbox{.}(2014)]%
        {hans2014did}
\bibfield{author}{\bibinfo{person}{Ronny Hans}, \bibinfo{person}{Ulrich Lampe}, \bibinfo{person}{Daniel Burgstahler}, \bibinfo{person}{Martin Hellwig}, {and} \bibinfo{person}{Ralf Steinmetz}.} \bibinfo{year}{2014}\natexlab{}.
\newblock \showarticletitle{Where did my battery go? quantifying the energy consumption of cloud gaming}. In \bibinfo{booktitle}{\emph{2014 IEEE International Conference on Mobile Services}}. IEEE, \bibinfo{pages}{63--67}.
\newblock


\bibitem[Inclezan and Pr{\'a}danos(2025)]%
        {inclezan2025environment}
\bibfield{author}{\bibinfo{person}{Daniela Inclezan} {and} \bibinfo{person}{Luis~I Pr{\'a}danos}.} \bibinfo{year}{2025}\natexlab{}.
\newblock \showarticletitle{Where is the environment in AI ethics? Integrating ecoliteracy into AI literacy}. In \bibinfo{booktitle}{\emph{International Conference on Artificial Intelligence in Education}}. Springer, \bibinfo{pages}{390--397}.
\newblock


\bibitem[Inie et~al\mbox{.}(2025)]%
        {inie2025co2stly}
\bibfield{author}{\bibinfo{person}{Nanna Inie}, \bibinfo{person}{Jeanette Falk}, {and} \bibinfo{person}{Raghavendra Selvan}.} \bibinfo{year}{2025}\natexlab{}.
\newblock \showarticletitle{How CO2STLY Is CHI? The Carbon Footprint of Generative AI in HCI Research and What We Should Do About It}. In \bibinfo{booktitle}{\emph{Proceedings of the 2025 CHI Conference on Human Factors in Computing Systems}}. \bibinfo{pages}{1--29}.
\newblock


\bibitem[Jegham et~al\mbox{.}(2025)]%
        {jegham2025hungry}
\bibfield{author}{\bibinfo{person}{Nidhal Jegham}, \bibinfo{person}{Marwen Abdelatti}, \bibinfo{person}{Lassad Elmoubarki}, {and} \bibinfo{person}{Abdeltawab Hendawi}.} \bibinfo{year}{2025}\natexlab{}.
\newblock \showarticletitle{How Hungry is AI? Benchmarking Energy, Water, and Carbon Footprint of LLM Inference}.
\newblock \bibinfo{journal}{\emph{arXiv preprint arXiv:2505.09598}} (\bibinfo{year}{2025}).
\newblock


\bibitem[Jordan et~al\mbox{.}(2021)]%
        {jordan2021poseblocks}
\bibfield{author}{\bibinfo{person}{Brian Jordan}, \bibinfo{person}{Nisha Devasia}, \bibinfo{person}{Jenna Hong}, \bibinfo{person}{Randi Williams}, {and} \bibinfo{person}{Cynthia Breazeal}.} \bibinfo{year}{2021}\natexlab{}.
\newblock \showarticletitle{PoseBlocks: A toolkit for creating (and dancing) with AI}. In \bibinfo{booktitle}{\emph{Proceedings of the AAAI Conference on Artificial Intelligence}}, Vol.~\bibinfo{volume}{35}. \bibinfo{pages}{15551--15559}.
\newblock


\bibitem[Kaijser and Kronsell(2014)]%
        {kaijser2014climate}
\bibfield{author}{\bibinfo{person}{Anna Kaijser} {and} \bibinfo{person}{Annica Kronsell}.} \bibinfo{year}{2014}\natexlab{}.
\newblock \showarticletitle{Climate change through the lens of intersectionality}.
\newblock \bibinfo{journal}{\emph{Environmental politics}} \bibinfo{volume}{23}, \bibinfo{number}{3} (\bibinfo{year}{2014}), \bibinfo{pages}{417--433}.
\newblock


\bibitem[Kewalramani et~al\mbox{.}(2021)]%
        {kewalramani2021using}
\bibfield{author}{\bibinfo{person}{Sarika Kewalramani}, \bibinfo{person}{Gillian Kidman}, {and} \bibinfo{person}{Ioanna Palaiologou}.} \bibinfo{year}{2021}\natexlab{}.
\newblock \showarticletitle{Using Artificial Intelligence (AI)-interfaced robotic toys in early childhood settings: a case for children’s inquiry literacy}.
\newblock \bibinfo{journal}{\emph{European Early Childhood Education Research Journal}} \bibinfo{volume}{29}, \bibinfo{number}{5} (\bibinfo{year}{2021}), \bibinfo{pages}{652--668}.
\newblock


\bibitem[Kissane(2020)]%
        {kissane2020integrating}
\bibfield{author}{\bibinfo{person}{B Kissane}.} \bibinfo{year}{2020}\natexlab{}.
\newblock \showarticletitle{Integrating technology into learning mathematics: the special place of the scientific calculator}. In \bibinfo{booktitle}{\emph{Journal of Physics: Conference Series}}, Vol.~\bibinfo{volume}{1581}. IOP Publishing, \bibinfo{pages}{012070}.
\newblock


\bibitem[Klein and D'Ignazio(2024)]%
        {klein2024data}
\bibfield{author}{\bibinfo{person}{Lauren Klein} {and} \bibinfo{person}{Catherine D'Ignazio}.} \bibinfo{year}{2024}\natexlab{}.
\newblock \showarticletitle{Data feminism for AI}. In \bibinfo{booktitle}{\emph{Proceedings of the 2024 ACM Conference on Fairness, Accountability, and Transparency}}. \bibinfo{pages}{100--112}.
\newblock


\bibitem[Lee et~al\mbox{.}(2021)]%
        {lee2021developing}
\bibfield{author}{\bibinfo{person}{Irene Lee}, \bibinfo{person}{Safinah Ali}, \bibinfo{person}{Helen Zhang}, \bibinfo{person}{Daniella DiPaola}, {and} \bibinfo{person}{Cynthia Breazeal}.} \bibinfo{year}{2021}\natexlab{}.
\newblock \showarticletitle{Developing middle school students' AI literacy}. In \bibinfo{booktitle}{\emph{Proceedings of the 52nd ACM technical symposium on computer science education}}. \bibinfo{pages}{191--197}.
\newblock


\bibitem[Lee et~al\mbox{.}(2020)]%
        {lee2020youth}
\bibfield{author}{\bibinfo{person}{Katharine Lee}, \bibinfo{person}{Nathalia Gjersoe}, \bibinfo{person}{Saffron O'neill}, {and} \bibinfo{person}{Julie Barnett}.} \bibinfo{year}{2020}\natexlab{}.
\newblock \showarticletitle{Youth perceptions of climate change: A narrative synthesis}.
\newblock \bibinfo{journal}{\emph{Wiley Interdisciplinary Reviews: Climate Change}} \bibinfo{volume}{11}, \bibinfo{number}{3} (\bibinfo{year}{2020}), \bibinfo{pages}{e641}.
\newblock


\bibitem[Lewis et~al\mbox{.}(2025)]%
        {lewis2025exploring}
\bibfield{author}{\bibinfo{person}{Aaleyah Lewis}, \bibinfo{person}{Aayushi Dangol}, \bibinfo{person}{Hyewon Suh}, \bibinfo{person}{Abbie Olszewski}, \bibinfo{person}{James Fogarty}, {and} \bibinfo{person}{Julie~A Kientz}.} \bibinfo{year}{2025}\natexlab{}.
\newblock \showarticletitle{Exploring ai-based support in speech-language pathology for culturally and linguistically diverse children}. In \bibinfo{booktitle}{\emph{Proceedings of the 2025 CHI Conference on Human Factors in Computing Systems}}. \bibinfo{pages}{1--19}.
\newblock


\bibitem[Li et~al\mbox{.}(2019)]%
        {li2019energy}
\bibfield{author}{\bibinfo{person}{Jingming Li}, \bibinfo{person}{Nianping Li}, \bibinfo{person}{Jinqing Peng}, \bibinfo{person}{Haijiao Cui}, {and} \bibinfo{person}{Zhibin Wu}.} \bibinfo{year}{2019}\natexlab{}.
\newblock \showarticletitle{Energy consumption of cryptocurrency mining: A study of electricity consumption in mining cryptocurrencies}.
\newblock \bibinfo{journal}{\emph{Energy}}  \bibinfo{volume}{168} (\bibinfo{year}{2019}), \bibinfo{pages}{160--168}.
\newblock


\bibitem[Li et~al\mbox{.}(2025)]%
        {li2025makingaithirstyuncovering}
\bibfield{author}{\bibinfo{person}{Pengfei Li}, \bibinfo{person}{Jianyi Yang}, \bibinfo{person}{Mohammad~A. Islam}, {and} \bibinfo{person}{Shaolei Ren}.} \bibinfo{year}{2025}\natexlab{}.
\newblock \bibinfo{title}{Making AI Less "Thirsty": Uncovering and Addressing the Secret Water Footprint of AI Models}.
\newblock
\newblock
\showeprint[arxiv]{2304.03271}~[cs.LG]
\urldef\tempurl%
\url{https://arxiv.org/abs/2304.03271}
\showURL{%
\tempurl}


\bibitem[Ligozat et~al\mbox{.}(2022)]%
        {ligozat2022unraveling}
\bibfield{author}{\bibinfo{person}{Anne-Laure Ligozat}, \bibinfo{person}{Julien Lef{\`e}vre}, \bibinfo{person}{Aur{\'e}lie Bugeau}, {and} \bibinfo{person}{Jacques Combaz}.} \bibinfo{year}{2022}\natexlab{}.
\newblock \showarticletitle{Unraveling the hidden environmental impacts of AI solutions for environment life cycle assessment of AI solutions}.
\newblock \bibinfo{journal}{\emph{Sustainability}} \bibinfo{volume}{14}, \bibinfo{number}{9} (\bibinfo{year}{2022}), \bibinfo{pages}{5172}.
\newblock


\bibitem[Long and Magerko(2020)]%
        {long2020ai}
\bibfield{author}{\bibinfo{person}{Duri Long} {and} \bibinfo{person}{Brian Magerko}.} \bibinfo{year}{2020}\natexlab{}.
\newblock \showarticletitle{What is AI literacy? Competencies and design considerations}. In \bibinfo{booktitle}{\emph{Proceedings of the 2020 CHI conference on human factors in computing systems}}. \bibinfo{pages}{1--16}.
\newblock


\bibitem[Luccioni and Hernandez-Garcia(2023)]%
        {luccioni2023counting}
\bibfield{author}{\bibinfo{person}{Alexandra~Sasha Luccioni} {and} \bibinfo{person}{Alex Hernandez-Garcia}.} \bibinfo{year}{2023}\natexlab{}.
\newblock \showarticletitle{Counting carbon: A survey of factors influencing the emissions of machine learning}.
\newblock \bibinfo{journal}{\emph{arXiv preprint arXiv:2302.08476}} (\bibinfo{year}{2023}).
\newblock


\bibitem[Madani et~al\mbox{.}(2017)]%
        {madani2017serious}
\bibfield{author}{\bibinfo{person}{Kaveh Madani}, \bibinfo{person}{Tyler~W Pierce}, {and} \bibinfo{person}{Ali Mirchi}.} \bibinfo{year}{2017}\natexlab{}.
\newblock \showarticletitle{Serious games on environmental management}.
\newblock \bibinfo{journal}{\emph{Sustainable cities and society}}  \bibinfo{volume}{29} (\bibinfo{year}{2017}), \bibinfo{pages}{1--11}.
\newblock


\bibitem[Maniates(2001)]%
        {maniates2001individualization}
\bibfield{author}{\bibinfo{person}{Michael~F Maniates}.} \bibinfo{year}{2001}\natexlab{}.
\newblock \showarticletitle{Individualization: Plant a tree, buy a bike, save the world?}
\newblock \bibinfo{journal}{\emph{Global environmental politics}} \bibinfo{volume}{1}, \bibinfo{number}{3} (\bibinfo{year}{2001}), \bibinfo{pages}{31--52}.
\newblock


\bibitem[McDonald et~al\mbox{.}(2019)]%
        {10.1145/3359174}
\bibfield{author}{\bibinfo{person}{Nora McDonald}, \bibinfo{person}{Sarita Schoenebeck}, {and} \bibinfo{person}{Andrea Forte}.} \bibinfo{year}{2019}\natexlab{}.
\newblock \showarticletitle{Reliability and Inter-rater Reliability in Qualitative Research: Norms and Guidelines for CSCW and HCI Practice}.
\newblock \bibinfo{journal}{\emph{Proc. ACM Hum.-Comput. Interact.}} \bibinfo{volume}{3}, \bibinfo{number}{CSCW}, Article \bibinfo{articleno}{72} (\bibinfo{date}{Nov.} \bibinfo{year}{2019}), \bibinfo{numpages}{23}~pages.
\newblock
\urldef\tempurl%
\url{https://doi.org/10.1145/3359174}
\showDOI{\tempurl}


\bibitem[Meadows(2008)]%
        {meadows2008thinking}
\bibfield{author}{\bibinfo{person}{Donella Meadows}.} \bibinfo{year}{2008}\natexlab{}.
\newblock \bibinfo{booktitle}{\emph{Thinking in systems: International bestseller}}.
\newblock \bibinfo{publisher}{chelsea green publishing}.
\newblock


\bibitem[Media(2025)]%
        {commonsense2025ai}
\bibfield{author}{\bibinfo{person}{Common~Sense Media}.} \bibinfo{year}{2025}\natexlab{}.
\newblock \bibinfo{booktitle}{\emph{New Report Shows Students Are Embracing Artificial Intelligence Despite Lack of Parent Awareness}}.
\newblock
\urldef\tempurl%
\url{https://www.commonsensemedia.org/press-releases/new-report-shows-students-are-embracing-artificial-intelligence-despite-lack-of-parent-awareness-and}
\showURL{%
\tempurl}
\newblock
\shownote{Accessed: 2025-01-26}.


\bibitem[Mott et~al\mbox{.}(2022)]%
        {mott2022robot}
\bibfield{author}{\bibinfo{person}{Terran Mott}, \bibinfo{person}{Alexandra Bejarano}, {and} \bibinfo{person}{Tom Williams}.} \bibinfo{year}{2022}\natexlab{}.
\newblock \showarticletitle{Robot co-design can help us engage child stakeholders in ethical reflection}. In \bibinfo{booktitle}{\emph{2022 17th ACM/IEEE International Conference on Human-Robot Interaction (HRI)}}. IEEE, \bibinfo{pages}{14--23}.
\newblock


\bibitem[Moulaison-Sandy and Thach(2025)]%
        {moulaison2025wicked}
\bibfield{author}{\bibinfo{person}{Heather Moulaison-Sandy} {and} \bibinfo{person}{Heather Thach}.} \bibinfo{year}{2025}\natexlab{}.
\newblock \showarticletitle{The Wicked Problem of AI: Information Avoidance, Uncomfortable Knowledge, and ChatGPT in Scholarly Communication}.
\newblock \bibinfo{journal}{\emph{Proceedings of the Association for Information Science and Technology}} \bibinfo{volume}{62}, \bibinfo{number}{1} (\bibinfo{year}{2025}), \bibinfo{pages}{1030--1035}.
\newblock


\bibitem[Moutaib et~al\mbox{.}(2020)]%
        {moutaib2020internet}
\bibfield{author}{\bibinfo{person}{Mohammed Moutaib}, \bibinfo{person}{Mohammed Fattah}, {and} \bibinfo{person}{Yousef Farhaoui}.} \bibinfo{year}{2020}\natexlab{}.
\newblock \showarticletitle{Internet of things: Energy consumption and data storage}.
\newblock \bibinfo{journal}{\emph{Procedia Computer Science}}  \bibinfo{volume}{175} (\bibinfo{year}{2020}), \bibinfo{pages}{609--614}.
\newblock


\bibitem[Nations(2025)]%
        {un2025water}
\bibfield{author}{\bibinfo{person}{United Nations}.} \bibinfo{year}{2025}\natexlab{}.
\newblock \bibinfo{title}{Artificial intelligence: How much energy does AI use?}
\newblock
\newblock
\urldef\tempurl%
\url{https://unric.org/en/artificial-intelligence-how-much-energy-does-ai-use/}
\showURL{%
\tempurl}
\newblock
\shownote{Accessed: 2026-01-26}.


\bibitem[Newman et~al\mbox{.}(2024)]%
        {newman2024want}
\bibfield{author}{\bibinfo{person}{Michele Newman}, \bibinfo{person}{Kaiwen Sun}, \bibinfo{person}{Ilena~B Dalla~Gasperina}, \bibinfo{person}{Grace~Y Shin}, \bibinfo{person}{Matthew~Kyle Pedraja}, \bibinfo{person}{Ritesh Kanchi}, \bibinfo{person}{Maia~B Song}, \bibinfo{person}{Rannie Li}, \bibinfo{person}{Jin~Ha Lee}, {and} \bibinfo{person}{Jason Yip}.} \bibinfo{year}{2024}\natexlab{}.
\newblock \showarticletitle{" I want it to talk like Darth Vader": Helping Children Construct Creative Self-Efficacy with Generative AI}. In \bibinfo{booktitle}{\emph{Proceedings of the CHI Conference on Human Factors in Computing Systems}}. \bibinfo{pages}{1--18}.
\newblock


\bibitem[Novaes(2025)]%
        {novaes2025enhancing}
\bibfield{author}{\bibinfo{person}{Andr{\'e}~Lucas Novaes}.} \bibinfo{year}{2025}\natexlab{}.
\newblock \showarticletitle{Enhancing sustainability education in higher education through simulation-based learning: integrating sustainable development goals}.
\newblock \bibinfo{journal}{\emph{International Journal of Sustainability in Higher Education}} (\bibinfo{year}{2025}).
\newblock


\bibitem[Noy and Zhang(2023)]%
        {noy2023experimental}
\bibfield{author}{\bibinfo{person}{Shakked Noy} {and} \bibinfo{person}{Whitney Zhang}.} \bibinfo{year}{2023}\natexlab{}.
\newblock \showarticletitle{Experimental evidence on the productivity effects of generative artificial intelligence}.
\newblock \bibinfo{journal}{\emph{Science}} \bibinfo{volume}{381}, \bibinfo{number}{6654} (\bibinfo{year}{2023}), \bibinfo{pages}{187--192}.
\newblock


\bibitem[O’Donnell and Crownhart(2025)]%
        {O’Donnell_Crownhart_2025}
\bibfield{author}{\bibinfo{person}{James O’Donnell} {and} \bibinfo{person}{Casey Crownhart}.} \bibinfo{year}{2025}\natexlab{}.
\newblock \bibinfo{title}{We did the math on AI’s energy footprint. here’s the story you haven’t heard.}
\newblock
\newblock
\urldef\tempurl%
\url{https://www.technologyreview.com/2025/05/20/1116327/ai-energy-usage-climate-footprint-big-tech/}
\showURL{%
\tempurl}


\bibitem[Pan et~al\mbox{.}(2025)]%
        {pan2025complexity}
\bibfield{author}{\bibinfo{person}{Liming Pan}, \bibinfo{person}{Chong-Yang Wang}, \bibinfo{person}{Fang Zhou}, {and} \bibinfo{person}{Linyuan L{\"u}}.} \bibinfo{year}{2025}\natexlab{}.
\newblock \showarticletitle{Complexity of social media in the era of generative AI}.
\newblock \bibinfo{journal}{\emph{National Science Review}} \bibinfo{volume}{12}, \bibinfo{number}{1} (\bibinfo{year}{2025}), \bibinfo{pages}{nwae323}.
\newblock


\bibitem[Patterson et~al\mbox{.}(2021)]%
        {patterson2021carbon}
\bibfield{author}{\bibinfo{person}{David Patterson}, \bibinfo{person}{Joseph Gonzalez}, \bibinfo{person}{Quoc Le}, \bibinfo{person}{Chen Liang}, \bibinfo{person}{Lluis-Miquel Munguia}, \bibinfo{person}{Daniel Rothchild}, \bibinfo{person}{David So}, \bibinfo{person}{Maud Texier}, {and} \bibinfo{person}{Jeff Dean}.} \bibinfo{year}{2021}\natexlab{}.
\newblock \showarticletitle{Carbon emissions and large neural network training}.
\newblock \bibinfo{journal}{\emph{arXiv preprint arXiv:2104.10350}} (\bibinfo{year}{2021}).
\newblock


\bibitem[Payne(2019)]%
        {payne2019ethics}
\bibfield{author}{\bibinfo{person}{Blakeley~H Payne}.} \bibinfo{year}{2019}\natexlab{}.
\newblock \showarticletitle{An ethics of artificial intelligence curriculum for middle school students}.
\newblock \bibinfo{journal}{\emph{MIT Media Lab Personal Robots Group. Retrieved Oct}}  \bibinfo{volume}{10} (\bibinfo{year}{2019}), \bibinfo{pages}{2019}.
\newblock


\bibitem[Peppler et~al\mbox{.}(2013)]%
        {peppler2013collaborative}
\bibfield{author}{\bibinfo{person}{Kylie Peppler}, \bibinfo{person}{Joshua~A Danish}, {and} \bibinfo{person}{David Phelps}.} \bibinfo{year}{2013}\natexlab{}.
\newblock \showarticletitle{Collaborative gaming: Teaching children about complex systems and collective behavior}.
\newblock \bibinfo{journal}{\emph{Simulation \& Gaming}} \bibinfo{volume}{44}, \bibinfo{number}{5} (\bibinfo{year}{2013}), \bibinfo{pages}{683--705}.
\newblock


\bibitem[Peretz(2025)]%
        {peretz2025integrating}
\bibfield{author}{\bibinfo{person}{Roee Peretz}.} \bibinfo{year}{2025}\natexlab{}.
\newblock \showarticletitle{Integrating Systems Thinking into Sustainability Education: An Overview with Educator-Focused Guidance}.
\newblock \bibinfo{journal}{\emph{Education Sciences}} \bibinfo{volume}{15}, \bibinfo{number}{12} (\bibinfo{year}{2025}), \bibinfo{pages}{1685}.
\newblock


\bibitem[Perez(2025)]%
        {perez2025growth}
\bibfield{author}{\bibinfo{person}{Sarah Perez}.} \bibinfo{year}{2025}\natexlab{}.
\newblock \bibinfo{title}{ChatGPT doubled its weekly active users in under 6 months, thanks to new releases}.
\newblock
\newblock
\urldef\tempurl%
\url{https://techcrunch.com/2025/03/06/chatgpt-doubled-its-weekly-active-users-in-under-6-months-thanks-to-new-releases/}
\showURL{%
\tempurl}
\newblock
\shownote{Accessed: 2026-01-26}.


\bibitem[Pinder et~al\mbox{.}(2018)]%
        {pinder2018digital}
\bibfield{author}{\bibinfo{person}{Charlie Pinder}, \bibinfo{person}{Jo Vermeulen}, \bibinfo{person}{Benjamin~R Cowan}, {and} \bibinfo{person}{Russell Beale}.} \bibinfo{year}{2018}\natexlab{}.
\newblock \showarticletitle{Digital behaviour change interventions to break and form habits}.
\newblock \bibinfo{journal}{\emph{ACM Transactions on Computer-Human Interaction (TOCHI)}} \bibinfo{volume}{25}, \bibinfo{number}{3} (\bibinfo{year}{2018}), \bibinfo{pages}{1--66}.
\newblock


\bibitem[Pousman and Stasko(2006)]%
        {pousman2006taxonomy}
\bibfield{author}{\bibinfo{person}{Zachary Pousman} {and} \bibinfo{person}{John Stasko}.} \bibinfo{year}{2006}\natexlab{}.
\newblock \showarticletitle{A taxonomy of ambient information systems: four patterns of design}. In \bibinfo{booktitle}{\emph{Proceedings of the working conference on Advanced visual interfaces}}. \bibinfo{pages}{67--74}.
\newblock


\bibitem[Reckien and Eisenack(2013)]%
        {reckien2013climate}
\bibfield{author}{\bibinfo{person}{Diana Reckien} {and} \bibinfo{person}{Klaus Eisenack}.} \bibinfo{year}{2013}\natexlab{}.
\newblock \showarticletitle{Climate change gaming on board and screen: A review}.
\newblock \bibinfo{journal}{\emph{Simulation \& Gaming}} \bibinfo{volume}{44}, \bibinfo{number}{2-3} (\bibinfo{year}{2013}), \bibinfo{pages}{253--271}.
\newblock


\bibitem[Resnick and Wilensky(1998)]%
        {resnick1998diving}
\bibfield{author}{\bibinfo{person}{Mitchel Resnick} {and} \bibinfo{person}{Uri Wilensky}.} \bibinfo{year}{1998}\natexlab{}.
\newblock \showarticletitle{Diving into complexity: Developing probabilistic decentralized thinking through role-playing activities}.
\newblock \bibinfo{journal}{\emph{The Journal of the Learning Sciences}} \bibinfo{volume}{7}, \bibinfo{number}{2} (\bibinfo{year}{1998}), \bibinfo{pages}{153--172}.
\newblock


\bibitem[Robinson and Ausubel(1983)]%
        {robinson1983game}
\bibfield{author}{\bibinfo{person}{Jennifer Robinson} {and} \bibinfo{person}{Jesse~H Ausubel}.} \bibinfo{year}{1983}\natexlab{}.
\newblock \showarticletitle{A game framework for scenario generation for the CO2 issue}.
\newblock \bibinfo{journal}{\emph{Simulation \& Games}} \bibinfo{volume}{14}, \bibinfo{number}{3} (\bibinfo{year}{1983}), \bibinfo{pages}{317--344}.
\newblock


\bibitem[Rogers(2018)]%
        {rogers2018coding}
\bibfield{author}{\bibinfo{person}{Richard Rogers}.} \bibinfo{year}{2018}\natexlab{}.
\newblock \showarticletitle{Coding and writing analytic memos on qualitative data: A review of Johnny Salda{\~n}a’s the coding manual for qualitative researchers}.
\newblock \bibinfo{journal}{\emph{The Qualitative Report}} \bibinfo{volume}{23}, \bibinfo{number}{4} (\bibinfo{year}{2018}), \bibinfo{pages}{889--892}.
\newblock


\bibitem[Satariano et~al\mbox{.}(2026)]%
        {satariano2026microsoftwater}
\bibfield{author}{\bibinfo{person}{Adam Satariano}, \bibinfo{person}{Paul Mozur}, {and} \bibinfo{person}{Karen Weise}.} \bibinfo{year}{2026}\natexlab{}.
\newblock \showarticletitle{Microsoft Pledged to Save Water. In the {A.I.} Era, It Expects Water Use to Soar}.
\newblock \bibinfo{journal}{\emph{The New York Times}} (\bibinfo{date}{27 Jan.} \bibinfo{year}{2026}).
\newblock
\urldef\tempurl%
\url{https://www.nytimes.com/2026/01/27/technology/microsoft-water-ai-data-centers.html}
\showURL{%
\tempurl}
\newblock
\shownote{Updated Jan.\ 27, 2026, 6:30 p.m.\ ET}.


\bibitem[Shapiro and Squire(2011)]%
        {shapiro2011games}
\bibfield{author}{\bibinfo{person}{R~Benjamin Shapiro} {and} \bibinfo{person}{Kurt~D Squire}.} \bibinfo{year}{2011}\natexlab{}.
\newblock \showarticletitle{Games for participatory science: A paradigm for game-based learning for promoting science literacy}.
\newblock \bibinfo{journal}{\emph{Educational Technology}} (\bibinfo{year}{2011}), \bibinfo{pages}{34--43}.
\newblock


\bibitem[Star and Ruhleder(2010)]%
        {star2010steps}
\bibfield{author}{\bibinfo{person}{Susan~Leigh Star} {and} \bibinfo{person}{Karen Ruhleder}.} \bibinfo{year}{2010}\natexlab{}.
\newblock \showarticletitle{Steps toward an Ecology of Infrastructure}.
\newblock \bibinfo{journal}{\emph{Revue d'anthropologie des connaissances}} \bibinfo{volume}{41}, \bibinfo{number}{1} (\bibinfo{year}{2010}), \bibinfo{pages}{114--161}.
\newblock


\bibitem[Suh et~al\mbox{.}(2024)]%
        {suh2024opportunities}
\bibfield{author}{\bibinfo{person}{Hyewon Suh}, \bibinfo{person}{Aayushi Dangol}, \bibinfo{person}{Hedda Meadan}, \bibinfo{person}{Carol~A Miller}, {and} \bibinfo{person}{Julie~A Kientz}.} \bibinfo{year}{2024}\natexlab{}.
\newblock \showarticletitle{Opportunities and challenges for AI-based support for speech-language pathologists}. In \bibinfo{booktitle}{\emph{Proceedings of the 3rd Annual Meeting of the Symposium on Human-Computer Interaction for Work}}. \bibinfo{pages}{1--14}.
\newblock


\bibitem[Tan and Mac(2026)]%
        {tan2026sky}
\bibfield{author}{\bibinfo{person}{Eli Tan} {and} \bibinfo{person}{Ryan Mac}.} \bibinfo{year}{2026}\natexlab{}.
\newblock \showarticletitle{Even the Sky May Not Be the Limit for {A.I.} Data Centers}.
\newblock \bibinfo{journal}{\emph{The New York Times}} (\bibinfo{date}{1 Jan.} \bibinfo{year}{2026}).
\newblock
\urldef\tempurl%
\url{https://www.nytimes.com/2026/01/01/technology/space-data-centers-ai.html}
\showURL{%
\tempurl}


\bibitem[Touretzky et~al\mbox{.}(2019)]%
        {touretzky2019envisioning}
\bibfield{author}{\bibinfo{person}{David Touretzky}, \bibinfo{person}{Christina Gardner-McCune}, \bibinfo{person}{Fred Martin}, {and} \bibinfo{person}{Deborah Seehorn}.} \bibinfo{year}{2019}\natexlab{}.
\newblock \showarticletitle{Envisioning AI for K-12: What should every child know about AI?}. In \bibinfo{booktitle}{\emph{Proceedings of the AAAI conference on artificial intelligence}}, Vol.~\bibinfo{volume}{33}. \bibinfo{pages}{9795--9799}.
\newblock


\bibitem[Vartiainen et~al\mbox{.}(2020)]%
        {vartiainen2020learning}
\bibfield{author}{\bibinfo{person}{Henriikka Vartiainen}, \bibinfo{person}{Matti Tedre}, {and} \bibinfo{person}{Teemu Valtonen}.} \bibinfo{year}{2020}\natexlab{}.
\newblock \showarticletitle{Learning Machine Learning with very Young Children: Who is Teaching Whom?}
\newblock \bibinfo{journal}{\emph{International journal of child-computer interaction}}  \bibinfo{volume}{25} (\bibinfo{year}{2020}), \bibinfo{pages}{100182}.
\newblock


\bibitem[Walsh et~al\mbox{.}(2013)]%
        {walsh2013facit}
\bibfield{author}{\bibinfo{person}{Greg Walsh}, \bibinfo{person}{Elizabeth Foss}, \bibinfo{person}{Jason Yip}, {and} \bibinfo{person}{Allison Druin}.} \bibinfo{year}{2013}\natexlab{}.
\newblock \showarticletitle{FACIT PD: a framework for analysis and creation of intergenerational techniques for participatory design}. In \bibinfo{booktitle}{\emph{proceedings of the SIGCHI Conference on Human Factors in Computing Systems}}. \bibinfo{pages}{2893--2902}.
\newblock


\bibitem[Weiser and Brown(1997)]%
        {weiser1997coming}
\bibfield{author}{\bibinfo{person}{Mark Weiser} {and} \bibinfo{person}{John~Seely Brown}.} \bibinfo{year}{1997}\natexlab{}.
\newblock \showarticletitle{The coming age of calm technology}.
\newblock In \bibinfo{booktitle}{\emph{Beyond calculation: The next fifty years of computing}}. \bibinfo{publisher}{Springer}, \bibinfo{pages}{75--85}.
\newblock


\bibitem[Wiek et~al\mbox{.}(2011)]%
        {wiek2011key}
\bibfield{author}{\bibinfo{person}{Arnim Wiek}, \bibinfo{person}{Lauren Withycombe}, {and} \bibinfo{person}{Charles~L Redman}.} \bibinfo{year}{2011}\natexlab{}.
\newblock \showarticletitle{Key competencies in sustainability: a reference framework for academic program development}.
\newblock \bibinfo{journal}{\emph{Sustainability science}} \bibinfo{volume}{6}, \bibinfo{number}{2} (\bibinfo{year}{2011}), \bibinfo{pages}{203--218}.
\newblock


\bibitem[Wilensky and Resnick(1999)]%
        {wilensky1999thinking}
\bibfield{author}{\bibinfo{person}{Uri Wilensky} {and} \bibinfo{person}{Mitchel Resnick}.} \bibinfo{year}{1999}\natexlab{}.
\newblock \showarticletitle{Thinking in levels: A dynamic systems approach to making sense of the world}.
\newblock \bibinfo{journal}{\emph{Journal of Science Education and technology}} \bibinfo{volume}{8}, \bibinfo{number}{1} (\bibinfo{year}{1999}), \bibinfo{pages}{3--19}.
\newblock


\bibitem[Williams et~al\mbox{.}(2023)]%
        {williams2023ai+}
\bibfield{author}{\bibinfo{person}{Randi Williams}, \bibinfo{person}{Safinah Ali}, \bibinfo{person}{Nisha Devasia}, \bibinfo{person}{Daniella DiPaola}, \bibinfo{person}{Jenna Hong}, \bibinfo{person}{Stephen~P Kaputsos}, \bibinfo{person}{Brian Jordan}, {and} \bibinfo{person}{Cynthia Breazeal}.} \bibinfo{year}{2023}\natexlab{}.
\newblock \showarticletitle{AI+ ethics curricula for middle school youth: Lessons learned from three project-based curricula}.
\newblock \bibinfo{journal}{\emph{International Journal of Artificial Intelligence in Education}} \bibinfo{volume}{33}, \bibinfo{number}{2} (\bibinfo{year}{2023}), \bibinfo{pages}{325--383}.
\newblock


\bibitem[Williams et~al\mbox{.}(2018)]%
        {williams2018my}
\bibfield{author}{\bibinfo{person}{Randi Williams}, \bibinfo{person}{Christian~V{\'a}zquez Machado}, \bibinfo{person}{Stefania Druga}, \bibinfo{person}{Cynthia Breazeal}, {and} \bibinfo{person}{Pattie Maes}.} \bibinfo{year}{2018}\natexlab{}.
\newblock \showarticletitle{" My doll says it's ok" a study of children's conformity to a talking doll}. In \bibinfo{booktitle}{\emph{Proceedings of the 17th ACM conference on interaction design and children}}. \bibinfo{pages}{625--631}.
\newblock


\bibitem[Williams et~al\mbox{.}(2019)]%
        {williams2019artificial}
\bibfield{author}{\bibinfo{person}{Randi Williams}, \bibinfo{person}{Hae~Won Park}, {and} \bibinfo{person}{Cynthia Breazeal}.} \bibinfo{year}{2019}\natexlab{}.
\newblock \showarticletitle{A is for artificial intelligence: the impact of artificial intelligence activities on young children's perceptions of robots}. In \bibinfo{booktitle}{\emph{Proceedings of the 2019 CHI conference on human factors in computing systems}}. \bibinfo{pages}{1--11}.
\newblock


\bibitem[Winner(2017)]%
        {winner2017artifacts}
\bibfield{author}{\bibinfo{person}{Langdon Winner}.} \bibinfo{year}{2017}\natexlab{}.
\newblock \showarticletitle{Do artifacts have politics?}
\newblock In \bibinfo{booktitle}{\emph{Computer ethics}}. \bibinfo{publisher}{Routledge}, \bibinfo{pages}{177--192}.
\newblock


\bibitem[Wise et~al\mbox{.}(2015)]%
        {wise2015kind}
\bibfield{author}{\bibinfo{person}{Alyssa~Friend Wise}, \bibinfo{person}{Alissa~Nicole Antle}, \bibinfo{person}{Jillian Warren}, \bibinfo{person}{Aaron May}, \bibinfo{person}{Min Fan}, {and} \bibinfo{person}{Anna Macaranas}.} \bibinfo{year}{2015}\natexlab{}.
\newblock \showarticletitle{What kind of world do you want to live in? Positive interdependence and collaborative processes in the tangible tabletop land-use planning game Youtopia}.
\newblock \bibinfo{publisher}{International Society of the Learning Sciences, Inc.[ISLS].}
\newblock


\bibitem[Wise et~al\mbox{.}(2021)]%
        {wise2021design}
\bibfield{author}{\bibinfo{person}{Alyssa~Friend Wise}, \bibinfo{person}{Alissa~N Antle}, {and} \bibinfo{person}{Jillian~L Warren}.} \bibinfo{year}{2021}\natexlab{}.
\newblock \showarticletitle{Design strategies for collaborative learning in tangible tabletops: Positive interdependence and reflective pauses}.
\newblock \bibinfo{journal}{\emph{Interacting with Computers}} \bibinfo{volume}{33}, \bibinfo{number}{3} (\bibinfo{year}{2021}), \bibinfo{pages}{271--294}.
\newblock


\bibitem[Wolfe et~al\mbox{.}(2024)]%
        {wolfe2024representation}
\bibfield{author}{\bibinfo{person}{Robert Wolfe}, \bibinfo{person}{Aayushi Dangol}, \bibinfo{person}{Bill Howe}, {and} \bibinfo{person}{Alexis Hiniker}.} \bibinfo{year}{2024}\natexlab{}.
\newblock \showarticletitle{Representation Bias of Adolescents in AI: A Bilingual, Bicultural Study}. In \bibinfo{booktitle}{\emph{Proceedings of the AAAI/ACM Conference on AI, Ethics, and Society}}, Vol.~\bibinfo{volume}{7}. \bibinfo{pages}{1621--1634}.
\newblock


\bibitem[Woodward et~al\mbox{.}(2018)]%
        {woodward2018using}
\bibfield{author}{\bibinfo{person}{Julia Woodward}, \bibinfo{person}{Zari McFadden}, \bibinfo{person}{Nicole Shiver}, \bibinfo{person}{Amir Ben-Hayon}, \bibinfo{person}{Jason~C Yip}, {and} \bibinfo{person}{Lisa Anthony}.} \bibinfo{year}{2018}\natexlab{}.
\newblock \showarticletitle{Using co-design to examine how children conceptualize intelligent interfaces}. In \bibinfo{booktitle}{\emph{Proceedings of the 2018 CHI conference on human factors in computing systems}}. \bibinfo{pages}{1--14}.
\newblock


\bibitem[Wu and Lee(2015)]%
        {wu2015climate}
\bibfield{author}{\bibinfo{person}{Jason~S Wu} {and} \bibinfo{person}{Joey~J Lee}.} \bibinfo{year}{2015}\natexlab{}.
\newblock \showarticletitle{Climate change games as tools for education and engagement}.
\newblock \bibinfo{journal}{\emph{Nature Climate Change}} \bibinfo{volume}{5}, \bibinfo{number}{5} (\bibinfo{year}{2015}), \bibinfo{pages}{413--418}.
\newblock


\bibitem[Wu(2023)]%
        {wu2023integrating}
\bibfield{author}{\bibinfo{person}{Yi Wu}.} \bibinfo{year}{2023}\natexlab{}.
\newblock \showarticletitle{Integrating generative AI in education: how ChatGPT brings challenges for future learning and teaching}.
\newblock \bibinfo{journal}{\emph{Journal of Advanced Research in Education}} \bibinfo{volume}{2}, \bibinfo{number}{4} (\bibinfo{year}{2023}), \bibinfo{pages}{6--10}.
\newblock


\bibitem[Yin(2013)]%
        {yin2013validity}
\bibfield{author}{\bibinfo{person}{Robert~K Yin}.} \bibinfo{year}{2013}\natexlab{}.
\newblock \showarticletitle{Validity and generalization in future case study evaluations}.
\newblock \bibinfo{journal}{\emph{Evaluation}} \bibinfo{volume}{19}, \bibinfo{number}{3} (\bibinfo{year}{2013}), \bibinfo{pages}{321--332}.
\newblock


\bibitem[Yip et~al\mbox{.}(2017)]%
        {10.1145/3025453.3025787}
\bibfield{author}{\bibinfo{person}{Jason~C. Yip}, \bibinfo{person}{Kiley Sobel}, \bibinfo{person}{Caroline Pitt}, \bibinfo{person}{Kung~Jin Lee}, \bibinfo{person}{Sijin Chen}, \bibinfo{person}{Kari Nasu}, {and} \bibinfo{person}{Laura~R. Pina}.} \bibinfo{year}{2017}\natexlab{}.
\newblock \showarticletitle{Examining Adult-Child Interactions in Intergenerational Participatory Design}. In \bibinfo{booktitle}{\emph{Proceedings of the 2017 CHI Conference on Human Factors in Computing Systems}} (Denver, Colorado, USA) \emph{(\bibinfo{series}{CHI '17})}. \bibinfo{publisher}{Association for Computing Machinery}, \bibinfo{address}{New York, NY, USA}, \bibinfo{pages}{5742–5754}.
\newblock
\showISBNx{9781450346559}
\urldef\tempurl%
\url{https://doi.org/10.1145/3025453.3025787}
\showDOI{\tempurl}


\bibitem[Zhang et~al\mbox{.}(2021)]%
        {zhang2021storydrawer}
\bibfield{author}{\bibinfo{person}{Chao Zhang}, \bibinfo{person}{Cheng Yao}, \bibinfo{person}{Jianhui Liu}, \bibinfo{person}{Zili Zhou}, \bibinfo{person}{Weilin Zhang}, \bibinfo{person}{Lijuan Liu}, \bibinfo{person}{Fangtian Ying}, \bibinfo{person}{Yijun Zhao}, {and} \bibinfo{person}{Guanyun Wang}.} \bibinfo{year}{2021}\natexlab{}.
\newblock \showarticletitle{StoryDrawer: A Co-Creative Agent Supporting Children's Storytelling through Collaborative Drawing}. In \bibinfo{booktitle}{\emph{Extended Abstracts of the 2021 CHI Conference on Human Factors in Computing Systems}}. \bibinfo{pages}{1--6}.
\newblock


\bibitem[Zhang et~al\mbox{.}(2023)]%
        {zhang2023integrating}
\bibfield{author}{\bibinfo{person}{Helen Zhang}, \bibinfo{person}{Irene Lee}, \bibinfo{person}{Safinah Ali}, \bibinfo{person}{Daniella DiPaola}, \bibinfo{person}{Yihong Cheng}, {and} \bibinfo{person}{Cynthia Breazeal}.} \bibinfo{year}{2023}\natexlab{}.
\newblock \showarticletitle{Integrating ethics and career futures with technical learning to promote AI literacy for middle school students: An exploratory study}.
\newblock \bibinfo{journal}{\emph{International Journal of Artificial Intelligence in Education}} \bibinfo{volume}{33}, \bibinfo{number}{2} (\bibinfo{year}{2023}), \bibinfo{pages}{290--324}.
\newblock


\bibitem[Zhang et~al\mbox{.}(2025)]%
        {zhang2025living}
\bibfield{author}{\bibinfo{person}{Zhihan Zhang}, \bibinfo{person}{Puvarin Thavikulwat}, \bibinfo{person}{Alexander~Le Metzger}, \bibinfo{person}{Yuxuan Mei}, \bibinfo{person}{Felix H{\"a}hnlein}, \bibinfo{person}{Zachary Englhardt}, \bibinfo{person}{Gregory~D Abowd}, \bibinfo{person}{Shwetak Patel}, \bibinfo{person}{Adriana Schulz}, \bibinfo{person}{Tingyu Cheng}, {et~al\mbox{.}}} \bibinfo{year}{2025}\natexlab{}.
\newblock \showarticletitle{Living sustainability: In-context interactive environmental impact communication}.
\newblock \bibinfo{journal}{\emph{Proceedings of the ACM on Interactive, Mobile, Wearable and Ubiquitous Technologies}} \bibinfo{volume}{9}, \bibinfo{number}{3} (\bibinfo{year}{2025}), \bibinfo{pages}{1--42}.
\newblock


\end{thebibliography}

\end{document}